\documentclass[fleqn,usenatbib]{mnras}

\usepackage{newtxtext,newtxmath}
\usepackage[T1]{fontenc}

\DeclareRobustCommand{\VAN}[3]{#2}
\let\VANthebibliography\thebibliography
\def\thebibliography{\DeclareRobustCommand{\VAN}[3]{##3}\VANthebibliography}

\usepackage{graphicx}	% Including figure files
\usepackage{amsmath}	% Advanced maths commands
\title[Accretion flows in the hard state of BHXBs]{Accretion flows in the hard state of black hole X-ray binaries: the effect of hot gas condensation}

\author[Wang et al.]{
Yilong Wang,$^{1,2}$ \thanks{wangyilong@nao.cas.cn} 
Bifang Liu,$^{1,2}$ \thanks{bfliu@nao.cas.cn}
Erlin Qiao$^{1,2}$
and Huaqing Cheng$^{1}$
\\
$^{1}$Key Laboratory of Space Astronomy and Technology, National Astronomical Observatories, Chinese Academy of Sciences, Beijing 100101, China\\
$^{2}$School of Astronomy and Space Science, University of Chinese Academy of Sciences, 19A Yuquan Road, Beijing 100049, China
}

\date{Accepted XXX. Received YYY; in original form ZZZ}

\pubyear{2023}
\begin{document}
\label{firstpage}
\pagerange{\pageref{firstpage}--\pageref{lastpage}}
\maketitle

\begin{abstract}
It is commonly believed that accretion discs are truncated and their inner regions are described by advection dominated accretion flows (ADAFs) in the hard spectral state of black hole X-ray binaries. However, the increasing occurrence of a relativistically blurred Fe K$\alpha$ line together with a hard continuum points to the existence of a thin disc located near the innermost stable circular orbit (ISCO). Assuming the accretion in the hard state is via an ADAF extending to near 100 Schwarzschild radii, which is supplied by either a stellar wind from a companion star or resulting from an evaporated disc, we study the possible condensation of the hot gas during its accretion towards the black hole. It is found that a small fraction of the ADAF condenses into a cold disc as a consequence of efficient radiative cooling at small distances, forming a disc-corona configuration near the ISCO. This takes place for low accretion rates corresponding to luminosities ranging from $\sim 10^{-3}$ to a few per cent of the Eddington luminosity. The coexistence of the weak inner disc and the dominant hot accretion flow provides a natural explanation of the broad K$\alpha$ line in the hard state. Detailed computations demonstrate that such accretion flows produce a hard X-ray spectrum accompanied by a weak disc component with a negative correlation between the 2-10 keV photon index and the Eddington ratio. The predicted spectrum of Cygnus X-1 and the correlation between the photon index and the Eddington ratio are in good agreement with observations.
\end{abstract}


%%%%%%%%%%%%%%%%%%%%%%%%%%%%%%%%%%%%%%%%%%%%%%%%%%

%%%%%%%%%%%%%%%%% BODY OF PAPER %%%%%%%%%%%%%%%%%%

\section{Introduction}

The coexistence of cold and hot accretion flows in the vicinity of black holes is a necessity for the theoretical interpretation of the observed spectral characteristics of black hole X-ray binaries (BHXBs) and active galactic nuclei (AGNs). The cold flow is generally described by the standard thin disc \citep{shakura1973, novikov1973} and the tenuous hot flow by the advection dominated accretion flow (ADAF) \citep{ichimaru1977, narayan1994, yuanf2014} or corona if it lies above/beneath a thin disc. In the past decades, various possible geometries of the two-phase, i.e., cold and hot, accretion flows around black holes have been explored \citep[for a review, see][]{poutanen2018}. Examples include an inner ADAF connected to a truncated thin disc \citep[e.g.][]{narayan1996, esinetal1997, esin1998, poutanen1997, poutanen2018, dove1997}, a thin disc sandwiched between slab-coronae \citep[e.g.][]{liang1977, haardt1991, haardt1993, svensson1994}, and patchy coronae lying on the disc surface \citep[e.g.][]{galeev1979, haardt1994, stern1995, svensson1996}, etc. These models, while incorporating the energy coupling between corona and disc, generally neglect the mass exchange between them. The disc evaporation model, which was originally established for dwarf novae \citep{meyer1994, liufk1995} and developed for black holes \citep[e.g.][]{liu1999,meyer2000a, meyer2000b, rozanska2000a,rozanska2000b,liu2002,meyer2003}, and its later extension, the corona condensation model \citep{liu2006, liu2007, meyer2007, taam2008, liu2011, qiao2012, qiao2013a, qiao2018b,cho2022}, include the consequential mass exchange associated with the energy balance between the two phases of the accretion flows \citep[for a review see][]{liu2022}. Such models provide a promising mechanism for the transition between the hot and cold accretion flows, and therefore, the formation of various structural accretion configurations.

It has been widely accepted that the accretion in BHXBs is dominated by the thin disc in the high/soft spectral state and the hot flow in the low/hard spectral state, and within the framework the continua can be well fit by a change in the configuration of the cold disc and the hot flow \citep[e.g.][]{esinetal1997, liu1999, meyer2007, liu2022}. However, the occurrence of a relativistically blurred Fe K$\alpha$ line in the hard spectrum \citep[e.g.][]{miller2006, miller2015, reis2010, parker2015, garcia2018, kara2019,buisson2019, ren2022, dong2022} is a challenge since it provides evidence for the coexistence of a cold disc near the ISCO with a dominant hot flow. We have demonstrated in our previous works \citep[e.g.][]{liu2007, meyer2007, taam2008} that in the hard state it is possible for the hot gas to condense and support a weak disc in the innermost region for a suitable range of accretion rates, taking into account the increasing coupling between electrons and ions at small distances. 
 
In this work, we reexamine the condensation of the hot accretion flow taking into account significant modifications to previous works in order to interpret the spectrum and constrain the occurrence of the broad Fe K$\alpha$ lines in the hard state of BHXBs. The main modifications include the following aspects. First, by adopting a self-similar solution for the hot accretion flow, we self-consistently calculate the radial dependence of the advection fraction of the viscous heating for a series of accretion rates, which is essential for determining the radiative efficiency of the hot flow as well as the emission spectrum. Secondly, soft photons from the central disc emission are included for the Comptonization in the corona/ADAF, and the local disc emission is calculated rather than approximated with a constant disc accretion rate. The illumination of the central coronal region of the disc is approximated as a lamp post. Finally, the critical accretion rate for the existence of an ADAF is calculated and the study of gas condensation is carried out only for accretion rates below this upper limit. 

In Section \ref{sec:model description} we describe in detail the corona condensation model and the improved calculational method. In Section \ref{sec:Numerical results} we present the numerical results based on the new method. In Section \ref{sec: comparison with observations} the theoretical predictions are compared with observations. Finally, a brief summary and discussion are given in Section \ref{sec:summary}. 

\section{Model Description}
\label{sec:model description}

The evaporation and condensation models take into account the energy coupling between the two-phase accretion flows through electron thermal conduction, the external Compton cooling of the corona by the disc soft photons and the reprocessing and reflection of the coronal irradiation. This coupling may result in either the evaporation of gas from the disc to the corona, described by the disc evaporation model \citep[e.g.][]{liu1999,meyer2000a, meyer2000b,rozanska2000a,rozanska2000b,liu2002,meyer2003}, or the condensation of hot coronal gas to the disc, as is the case in the corona condensation model \citep{liu2006, liu2007, meyer2007, taam2008, liu2011, qiao2012, qiao2013a, qiao2018b,cho2022}. An equilibrium is eventually established in the two-phase accretion flow and the final steady geometry of accretion in general depends on the mass transfer rate and its mode, i.e., via  Roche lobe overflow (RLOF) or stellar wind.

In the case of accretion via RLOF as in low-mass black hole X-ray binaries (LMXBs), gas transferred from the secondary is confined to the orbital plane and forms a thin disc in the outer region, and a corona, which is fed by the evaporation of the disc. The geometry of such accretion flows is either a disc extending to the ISCO with a hot corona lying above its surface, or an ADAF connected to an evaporation-induced truncated disc when the gas supply rate is too low to compete with the evaporation \citep[e.g.][]{liu1999, meyer2000a}. This gives rise to the observed high/soft spectral state or the low/hard spectral state, respectively.

In the case of wind-fed accretion as in high-mass X-ray binaries (HMXBs), the hot stellar wind of the massive companion captured by the black hole directly forms an ADAF-like hot accretion flow, as long as the gas supply rate is below the critical accretion rate for an ADAF to exist. If the gas supply rate exceeds this limit, the captured stellar wind would collapse into a thin disc, which is likely sandwiched in a residual corona. This also gives rise to the hard and soft spectral states, respectively.

In the hard state, whether gas is supplied by the RLOF or the stellar wind, the nature of accretion in the inner region is essentially similar. Even if a thin disc forms as in the case of RLOF, the evaporation leads to a change in the accretion configuration with the transition of the thin disc into an ADAF at distances of about 200 Schwarzschild radii ($R_{\rm S}$) \citep[e.g.][]{meyer2000b, liu2002}. Here, the accretion in the inner region is not significantly affected by the form of the outer accretion flow (via ADAF or thin disc). In the following, we focus on the accretion flows in the innermost region ($R<100R_S$) and reinvestigate the conditions under which condensation of the hot accretion flow takes place in the hard state.

\subsection{Physics of condensation}

Hot gas condensation is a consequence of energy balance in the accretion flows, which usually occurs in the inner region where radiation is important in cooling the hot gas. At high accretion rates, the external Compton cooling is very efficient, leading to strong condensation and a very weak corona. On the other hand, at medium or low accretion rates, condensation of the corona/ADAF is no longer strong but still possible in the inner region \citep[e.g.][]{liu2006, liu2007, meyer2007}.

In this work we investigate the steady-state structure of the accretion flow at low accretion rates, which is initially an ADAF-like hot flow at about $100 R_{\rm S}$, supplied by either the stellar wind or the evaporated disc. When the hot gas steadily flows toward the black hole, the density increases and both the Coulomb coupling and radiation become efficient. If the accretion rate is not very low, a part of the hot flow can condense and form a cold disc, with the remainder of the hot flow lying above the disc. Thus, the geometry of the accretion flow can be described as an outer ADAF-like region plus an inner disc-corona configuration. The hot and cold accretion flows mutually interact through mechanisms including electron thermal conduction, Comptonization of the disc soft photons in the corona, the reprocessing and reflection of the coronal irradiation in the disc, and gas condensation from the corona to the disc. Such a model differs from traditional disc-corona models \citep[e.g.][]{galeev1979, haardt1991, esinetal1997} in that the gas exchange between the cold disc and hot corona is included.

We use a semi-analytical corona condensation model with an improved numerical method to study the interaction between the corona and the disc. Specifically, the interaction is treated in a vertically stratified method, which divides the accretion flow into three parts in the vertical direction, namely a hot corona, a cold disc, and a transition layer in between, which are tightly coupled \citep[e.g.][]{liu2007, meyer2007}. The corona is physically similar to a pure ADAF except for its interaction with the underlying disc, where the Compton cooling of the disc soft photons can efficiently increase its radiative efficiency, which is reflected by the general decrease of the advection fraction of the viscously dissipated energy. The transition layer is a cooler and denser thin layer of the corona with coupled ion and electron temperatures. This is distinct from the major part of the corona which is hot and geometrically thick, with decoupled ion and electron temperatures due to the inefficient Coulomb collisions caused by low particle densities. The dominant cooling and heating mechanisms in the transition layer are bremsstrahlung and the electron thermal conduction from the hot corona, respectively. When the heating by thermal conduction exceeds the bremsstrahlung cooling, a certain amount of cool gas is heated up into the corona, i.e., evaporated, until an energy equilibrium is established. In the opposite case when the thermal conduction heating is less efficient as compared to the bremsstrahlung cooling, hot gas is over-cooled and condensates into the cold disc. Essentially, the energy balance and resultant evaporation or condensation depend on the accretion rate, which are described in detail in the following subsections.

\subsection{The corona}
\label{subsec: the corona}

The corona can be described, in form,  by  the self-similar solution of an ADAF, with the electron number density $n_{\rm e}$,  the pressure $p$, the viscous heating rate $q^+$, the electron scattering optical depth $\tau_{\rm es}$, and the magnetic field strength $B$ expressed as \citep{narayan1995b},
\begin{gather}
    n_{\rm e} = 2.0 \times 10^{19} \alpha^{-1} c_1^{-1} c_3^{-1/2} m^{-1} \dot{m}_{\rm cor} r^{-3/2}\ {\rm cm^{-3}}, \notag \\ 
    p = 1.71 \times 10^{16} \alpha^{-1} c_1^{-1} c_3^{1/2} m^{-1} \dot{m}_{\rm cor} r^{-5/2} \ {\rm g\ cm^{-1}\ s^{-2}}, \notag \\
    q^+ = 1.84 \times 10^{21} \epsilon' c_3^{1/2} m^{-2} \dot{m}_{\rm cor} r^{-4} \ {\rm ergs\ cm^{-3}\ s^{-1}}, \notag \\ 
    \tau_{\rm es} = 12.4 \alpha^{-1} c_1^{-1} \dot{m}_{\rm cor} r^{-1/2},\notag \\
    B = 6.55 \times 10^{8} \alpha^{-1/2} \left(1 - \beta \right)^{1/2} c_1^{-1/2} c_3^{1/4} m^{-1/2} \dot{m}_{\rm cor}^{1/2} r^{-5/4}\ {\rm G}, 
    \label{eq:self-similar solutions}
\end{gather}
where $m$ is the black hole mass scaled by the solar mass $M_{\odot}$, $\dot{m}_{\rm cor}$ is the mass accretion rate in the corona scaled by the Eddington accretion rate $\dot{M}_{\rm Edd} \approx 1.39 \times 10^{18} m \ {\rm g\ s^{-1}}$, $r$ is the radius scaled by $R_{\rm S} = 2.95 \times 10^5 m \ {\rm cm}$, $\alpha$ is the viscosity parameter, and $\beta$ is the magnetic parameter defined as $1-\beta \equiv p_{\rm m}/p$, where $p_{\rm m}=B^2/ 24\pi$  is the magnetic pressure, $p$ the sum of the gas pressure and magnetic pressure. We take $\beta = 0.95$ in all of our calculations as suggested by the numerical simulations of \citet{hawley2001}. The coefficients in the self-similar solution are derived as,
\begin{gather}
    c_1 = \frac{5+2\epsilon'}{3\alpha^2}g(\alpha,\epsilon')\approx \frac{3}{5+2\epsilon'}, \notag  \\
    c_3 = \frac{2}{3}c_1, \notag \\
    g(\alpha,\epsilon')=\left[1+\frac{18\alpha^2}{(5+2\epsilon')^2}\right]^{1/2}-1\approx \frac{9\alpha^2}{(5+2\epsilon')^2}, \notag \\
    \epsilon' = \frac{\epsilon}{f}, \notag \\
    \epsilon = \frac{5/3-\gamma}{\gamma-1},\notag \\
    \gamma=\frac{8-3\beta}{6-3\beta},
    \label{eq:coefficients}
\end{gather}
where $f$ is the advection fraction of the viscous heating, which is to be determined by  combining the energy equations respectively for ions and electrons, and the equation of state. The approximate relations in the above equations are valid when $\alpha^2 \ll 1$, which almost always applies in our calculations. 

We point out that the above solutions describing the coronal parameters differ from a typical ADAF as they depend on the advection fraction of the viscously dissipated energy ($f$) via $c_1$, $c_3$ and $\epsilon'$. In the ADAF, most of the viscous heating is advected, and thus, by setting $f\sim1$, the pressure, density, and ion temperature can be determined approximately for given $m$, $\dot m$, $\alpha$ and $\beta$, with the electron temperature determined by the energy balance between collisional heating and radiative cooling. In the corona, the inverse Compton scattering off the soft photons from the disc and the thermal conduction from the corona to the transition layer lead to more efficient cooling and collisional heating, which can result in a small value of $f$. Therefore, $f$ should be self-consistently calculated, for which we need not only to solve the complete set of equations describing the ADAF, but also to combine the equations describing the corona, the transition layer, and the disc, as they are coupled through radiation and gas condensation. This is one of the major improvements made in this work, while an estimated value, $f=0.05$, was fixed in our previous works.

In the corona, the ions are directly heated by viscosity at a rate of $q^+$, and the energy they receive is partially transferred to the electrons via Coulomb collisions, with the remaining part of the energy advected. Thus, the energy balance equation of the ions in the corona/ADAF is, by the definition of $f$,
\begin{equation}
    q^+ = q_{\rm ie} + f q^+,
    \label{eq:ion energy}
\end{equation}
where $q_{\rm ie}$ refers to the volume energy transfer rate from ions to electrons via Coulomb collisions, which is expressed as \citep{stepney1983,liu2002},
\begin{gather}
    q_{\rm ie} = 3.59 \times 10^{-32} n_{\rm e}n_{\rm i}(T_{\rm i}-T_{\rm e})\left(\frac{1+T_*^{1/2}}{T_*^{3/2}}\right) , \notag \\
    T_* = \frac{k T_{\rm e}}{m_{\rm e}c^2}\left(1+\frac{m_{\rm e}T_{\rm i}}{m_{\rm p}T_{\rm e}}\right),
\end{gather}
with $k$ being the Boltzmann constant, $c$ the light speed, $m_{\rm e}$ the electron mass, $m_{\rm p}$ the proton mass, $T_{\rm e}$ the electron temperature, and $T_{\rm i}$ the ion temperature. The ion number density $n_{\rm i}$ is related to $n_{\rm e}$ by $n_{\rm i}=n_{\rm e}/1.077$ for an assumed chemical abundance of hydrogen mass fraction $X=0.75$ and helium mass fraction $Y = 0.25$. For this chemical abundance, the ion temperature $T_{\rm i}$ and electron temperature $T_{\rm e}$ can be related by the equation of state \citep{narayan1995b},
\begin{equation}
    T_{\rm i} + 1.08T_{\rm e} = 6.66 \times 10^{12} \beta c_3 r^{-1}\ {\rm K}.
    \label{eq:EOS}
\end{equation}

For the heating of electrons, we consider only Coulomb collisions and neglect their direct heating by viscosity. The cooling of electrons, on the other hand, is much more complicated as it involves multiple radiative processes in the corona, as well as the thermal conduction from the hot corona to the cooler transition layer. Therefore, the energy balance equation of the electrons is,
\begin{equation}
    q_{\rm ie} = q_{\rm rad} + \Delta F_{\rm c}/H,
    \label{eq:electron energy}
\end{equation}
where $H = (2.5 c_3)^{1/2}r R_{\rm S}$ \citep{narayan1995b} is the vertical scale height  of the corona, $\Delta F_{\rm c} \approx k_0 T_{\rm e}^{7/2}/H$ \citep[$k_0 = 10^{-6}{\rm \,ergs\,s^{-1}\,cm^{-1}\,K^{-7/2}}$,][]{spitzer1962} refers to the thermal conduction flux from the corona to the transition layer, and $q_{\rm rad}$ is the volume radiative cooling rate of the electrons, which is given by 
\begin{equation}
    q_{\rm rad} = q_{\rm br} + q_{\rm syn} + q_{\rm self,cmp} + q_{\rm ex,cmp},
\end{equation}
where $q_{\rm br}$, $q_{\rm syn}$ and $q_{\rm self,cmp}$ are the bremsstrahlung cooling rate, synchrotron cooling rate and the corresponding self-Compton cooling rate respectively, which are all functions of $n_{\rm e}$, $T_{\rm e}$ and $H$ \citep{narayan1995b, manmoto1997}. $q_{\rm ex,cmp}$ refers to the cooling caused by the Comptonization of the external soft photons from the underlying disc, which are produced by both the viscous process of the disc and the reprocessing of the irradiation photons from the corona in the disc.

The radiative coupling and gas exchange between the disc and the corona complicate the calculation of soft photons for Compton scattering, which is improved in this paper. In the previous works \citep[e.g.][]{qiao2012,qiao2013a,liu2015, qiao2018b, taam2018}, the soft photons from the radiation associated with the local viscous dissipation of the disc were approximated by the standard disc emission with an accretion rate accumulated from the condensation at all distances. This approximation over-estimated the soft photon flux of the outer disc where the accretion rate supplied by condensation is small as compared to the total accumulated one, which led to higher condensation rate and consequently a much softer spectrum, as is discussed in Section \ref{sec:Numerical results}. In this work we take the local disc accretion rate $\dot m_{\rm disc} (r)$ at each radius, for which additional iterations are necessary in the computation. In addition, we consider the contribution of photons emitted from the central disc to the soft photons, which can overwhelm the local emission at large distances. However, for the Comptonization in the innermost corona, this central disc emission term should be dropped in the computation as it is dominantly contributed by the local emission which has been included in the local soft photons. The third source of soft photons for Comptonization is the irradiation of the disc by the corona, which is reprocessed in the disc as soft photons except for a small reflected fraction. We use a lamp-post model which simplifies the corona as a point source above the black hole, at a height of $H_{\rm s} = 10R_{\rm S}$ (see Section \ref{subsec: lamp-post} for a discussion on the lamp-post approximation). Thus, the total soft photon flux for Comptonization in the corona at any distance, $F_{\rm soft}(R)$, which is composed of the flux of local disc emission, $F_{\rm local}(R)$, the flux of central disc emission, $F_{\rm central}(R)$, and the reprocessed coronal irradiation, $F_{\rm irr}(R)$, is now expressed as,
\begin{gather}
        F_{\rm soft}(R) = {\rm max}\left[F_{\rm local}(R), F_{\rm central}(R)\right] + F_{\rm irr}(R), \notag \\
        F_{\rm local}(R) = \frac{3GM\dot{M}_{\rm disc} (R)}{8\pi R^3}\left(1 - \sqrt{\frac{3R_{\rm S}}{R}}\right), \notag \\
        F_{\rm central}(R)= \left(1 - 1/\sqrt{2}\right)\frac{L_{\rm disc}}{8\pi R^2}, \notag \\
        F_{\rm irr} (R)= \frac{(1-a)L_{\rm cor}}{8\pi}\frac{H_{\rm s}}{(R^2 + H^2_{\rm s})^{3/2}},
        \label{eq:soft photon flux}
\end{gather}
where $M = m M_{\odot}$, $\dot{M}_{\rm disc} = \dot{m}_{\rm disc} \dot{M}_{\rm Edd}$, $R = r R_{\rm S}$. $G$ is the gravitational constant, and $a$ is the albedo of the coronal irradiation. $a = 0.15$ is adopted in our calculations as suggested by numerical simulations \citep{magdziarz1995}. $F_{\rm central}(R)$ is averaged along the corona height, leading to the factor $1- 1/\sqrt{2}$. A radiative transfer factor, $e^{-\tau (R)}$, is also added to $F_{\rm central}(R)$ when the radial optical depth is close to 1. For simplicity we have neglected $F_{\rm local}$ when $F_{\rm local} < F_{\rm central}$. The luminosity of the disc ($L_{\rm disc}$) and the luminosity of the corona ($L_{\rm cor}$) are given as,
\begin{gather}
    L_{\rm disc} = 2\int_{R_{\rm in}}^{R_{\rm cnd}} \left(F_{\rm local}+ F_{\rm irr}\right) 2\pi R dR, \notag \\
   L_{\rm cor} = 2\int_{R_{\rm in}}^{R_{\rm out}} q_{\rm rad} H 2\pi R dR.
\end{gather}
Here, $R_{\rm cnd}$ is the radius where the inflowing hot gas starts to condense and hence corresponds to the outer radius of the condensation-fed disc, and $R_{\rm in}$ is the radius of the inner boundary of the accretion flow. $R_{\rm out}$ is the outermost radius of the accretion flow from which our calculations start, and it has only limited influence on the numerical results as long as $200 R_{\rm S} \gtrsim R_{\rm out} > R_{\rm cnd}$, since the radiation of the outer part of the accretion flow is insignificant. The upper limit $200 R_{\rm S}$ is adopted because the RLOF-fed accretion disc in the hard state of LMXBs is fully evaporated at $\sim 200 R_{\rm S}$ \citep[e.g.][]{meyer2000b} and hence the condensation model is applicable, although such a constraint is not necessary for the wind-fed accretion systems. In this work $R_{\rm out} = 100 R_{\rm S}$ is taken as an example. Thus, with $F_{\rm soft}$ determined at each radius $R$, the energy density of the disc soft photons $u(R)$ and the corresponding external Compton cooling rate $q_{\rm ex,cmp}$ can be expressed as,
\begin{gather}
    u(R) = \frac{2}{c} F_{\rm soft},  \notag \\
    q_{\rm ex,cmp} = \frac{4kT_{\rm e}}{m_{\rm e}c^2} n_{\rm e} \sigma_{\rm T} c u,
\end{gather}
with $\sigma_{\rm T}$ being the Thomson scattering optical depth.

Given the basic parameters, $\alpha$, $\beta$, $a$, $m$, and $\dot m$, the corona structure (described by $T_{\rm i}$, $T_{\rm e}$, $n_{\rm e}$, $f$) and radiation features can be determined by the above equations, provided that the accretion rates in the corona and in the disc are known ($\dot m_{\rm disc}(r)+\dot m_{\rm cor}(r)=\dot m$). Thus, it is necessary to combine the disc, the transition layer, and the corona for a detailed study of gas condensation to the disc as follows.

\subsection{From ADAF to corona}

For the sake of clarity it is worth re-emphasizing the differences between the ADAF and the corona described above. The self-similar solution given by \citet{narayan1994,narayan1995a,narayan1995b} has three solution branches, namely the ADAF branch, the unstable branch, and the cooling dominated branch. We take the ADAF branch to describe the optically thin hot accretion flow, which is essentially similar to the ADAF. In fact, it is exactly the ADAF when the hot gas supply rate is so low that no condensation occurs. However, there are differences at higher accretion rates because a part of the hot gas condenses to a thin disc as a consequence of the extra vertical thermal conduction and the external Compton cooling.

Condensation takes place in the inner region where the hot gas is compressed into a small volume, thereby increasing the efficiency of radiative cooling. When a thin disc can be steadily supplied by the gas condensation, the hot accretion flow, which is now called a corona, emits efficiently by the additional Compton scattering, which results in a much smaller fraction of advection energy. As compared to the inner corona, the outer hot accretion flow is akin to an ADAF, because at radii beyond the condensation radius the only influence due to the existence of a cold disc is the extra Compton cooling caused by soft photons from the inner disc. In the previous works \citep[e.g.][]{qiao2012,qiao2013a,liu2015, qiao2018b, taam2018}, the contribution of the outer part of the hot flow was neglected, which is a reasonable simplification when $R_{\rm cnd}$ is large, since most of the radiation is from the corona. In this work, we consider both the inner disc-corona configuration and the outer hot accretion flow. The equations and related calculations of the outer hot flow are nearly the same as those of the corona, except that we replace $\dot{m}_{\rm cor}$ with the total mass accretion rate $\dot{m}$ (in unit of $\dot{M}_{\rm Edd}$), $\Delta F_{\rm c}$ in equation (\ref{eq:electron energy}) and $F_{\rm irr}$ are set to zero, and, implied by its definition, $F_{\rm local} = 0$.

\subsection{The transition layer}
\label{s:trans}

Following the previous works, we consider a transition layer between the corona and the disc, in which the difference between the bremsstrahlung cooling and the heating of electron thermal conduction leads to either the condensation of coronal gas to the disc or the evaporation of the disc matter to the corona. In the low/hard state of BHXBs, only the former takes place. From the energy balance of electrons in the transition layer, the condensation rate per unit area ($\dot m_{\rm z}$) is given as \citep[e.g.][]{liu2007,meyer2007},
\begin{gather}
    \dot{m}_{\rm z} = \frac{\gamma-1}{\gamma}\beta \frac{\Delta F_{\rm c}}{kT_{\rm i}/\mu_{\rm i}m_{\rm p}}\left(1-\sqrt{C}\right), \notag \\
    C \equiv k_0b\left(\frac{\beta^2p^2}{\pi k^2}\right)\left(\frac{T_{\rm cpl}}{\Delta F_{\rm c}}\right)^2,
    \label{eq:condensation rate}
\end{gather}
where $\mu_{\rm i} = 4/(1+3X) \approx 1.23$ is the effective molecular weight of the ions and $T_{\rm cpl}$ is the coupling temperature in the transition layer determined by the energy balance equation of the ions:
\begin{equation}
    q^+ + q_{\rm c} =  \frac{2}{\sqrt{\pi}}q_{\rm ie}.
\end{equation}
$q_{\rm c} = \frac{f}{1-\beta}q^+$ \citep{esin1997} is the compressive heating rate, which is all transferred to electrons since there is no change in the coupling temperature and the internal energy. The factor $\frac{2}{\sqrt{\pi}}$ is included to correct the difference between the densities of the corona and the transition layer.  

\subsection{The disc}

When the gas supply is sufficient for the condensation to take place, an inner disc exists within a critical condensation radius $R_{\rm cnd}$, which is determined by setting $C = 1$ in equation (\ref{eq:condensation rate}). From $R_{\rm cnd}$ inward, $\dot{m}_{\rm z} <0$, indicating that the ADAF-like hot accretion flow partially condensates into the disc in this area, and therefore the accretion flow is in the disc-corona configuration. While in the outer region, evaporation occurs ($\dot{m}_{\rm z} >0$) if a disc exists, which eventually evacuates the cold disc and a steady hot accretion flow is preserved. $R_{\rm cnd}$ is anticipated to depend on the mass-supply rate, with a larger disc at a higher supply rate. The accretion rate in the disc, $\dot{m}_{\rm disc}(R)$, is given by integration of $\dot{m}_{\rm z}$ over the area between $R$ and $R_{\rm cnd}$:
\begin{equation}
    \dot{m}_{\rm disc}(R) = \int_{R}^{R_{\rm cnd}}4\pi R\frac{\dot{m}_{\rm z}}{\dot{M}_{\rm Edd}}dR,
\end{equation}
and the coronal accretion rate $\dot{m}_{\rm cor}(R)$ is given by mass conservation as,
\begin{equation}
    \dot{m}_{\rm cor}(R) = \dot{m} - \dot{m}_{\rm disc}(R).
\end{equation}

\subsection{The improved numerical calculation method}
\label{subsec: calculation method}

In this work, we improve on the previous numerical calculation method of the corona condensation model by iterating to calculate self-consistent solutions of the above equations. For each calculation, we fix $\alpha$, $\beta$, $a$, $m$, and $\dot{m}$ as basic input parameters and start with presumed values of $L_{\rm cor}$ and $L_{\rm disc}$. From the outer boundary inward, at each radius $R$ we use the Levenberg-Marquardt algorithm to numerically solve for $T_{\rm i}(R)$, $T_{\rm e}(R)$, $\dot{m}_{\rm z}(R)$, from equations (\ref{eq:ion energy}), (\ref{eq:EOS}), and (\ref{eq:condensation rate}), supplemented by the relevant expressions for the quantities involved as listed above. Most importantly, the advection factor, $f(R)$, is self-consistently determined by iteration at each distance. In the process, $\dot{m}_{\rm disc}$, $\dot{m}_{\rm cor}$, $L_{\rm disc}$, $L_{\rm cor}$ and other physical quantities are also derived. The deviations between the presumed and derived values of $L_{\rm cor}$ and $L_{\rm disc}$ are checked and the calculation is repeated by assigning updated values of $L_{\rm cor}$ and $L_{\rm disc}$. The iterative calculation continues until such deviations are less than $1\%$. With the determination of the structure of the accretion flows, we compute the multi-color black body radiation from the disc, the bremsstrahlung, synchrotron, and Comptonization radiations from the corona and the ADAF, where the inverse Compton scatterings are calculated from Monte Carlo simulations \citep{pozdniakov1977,qiao2012}.

\section{Numerical results}
\label{sec:Numerical results}

\begin{figure}
    \centering
    \includegraphics[width=\linewidth]{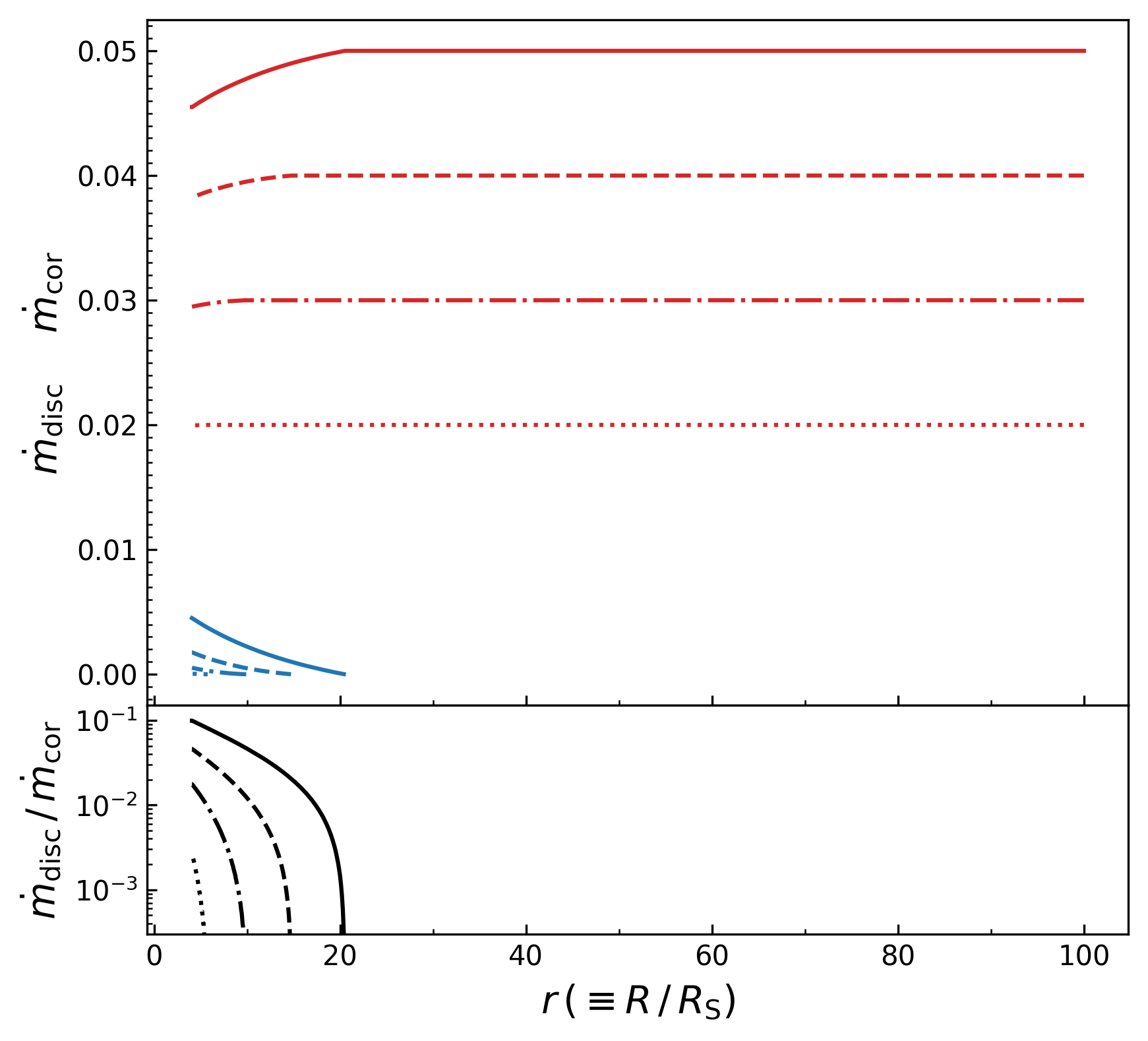}
    \caption{Upper panel: the radial distribution of accretion rates in the hot accretion flows (red lines) and in the discs (blue lines) with mass-supply rate $\dot{m} = $ 0.05 (solid lines), 0.04 (dashed lines), 0.03 (dotted-dashed lines), 0.02 (dotted lines). Lower panel: the ratio of the disc accretion rate to the coronal accretion rate. The same line styles as the upper panel are adopted.}
    \label{fig:mdot}
\end{figure}

\begin{figure}
    \centering
    \includegraphics[width=\linewidth]{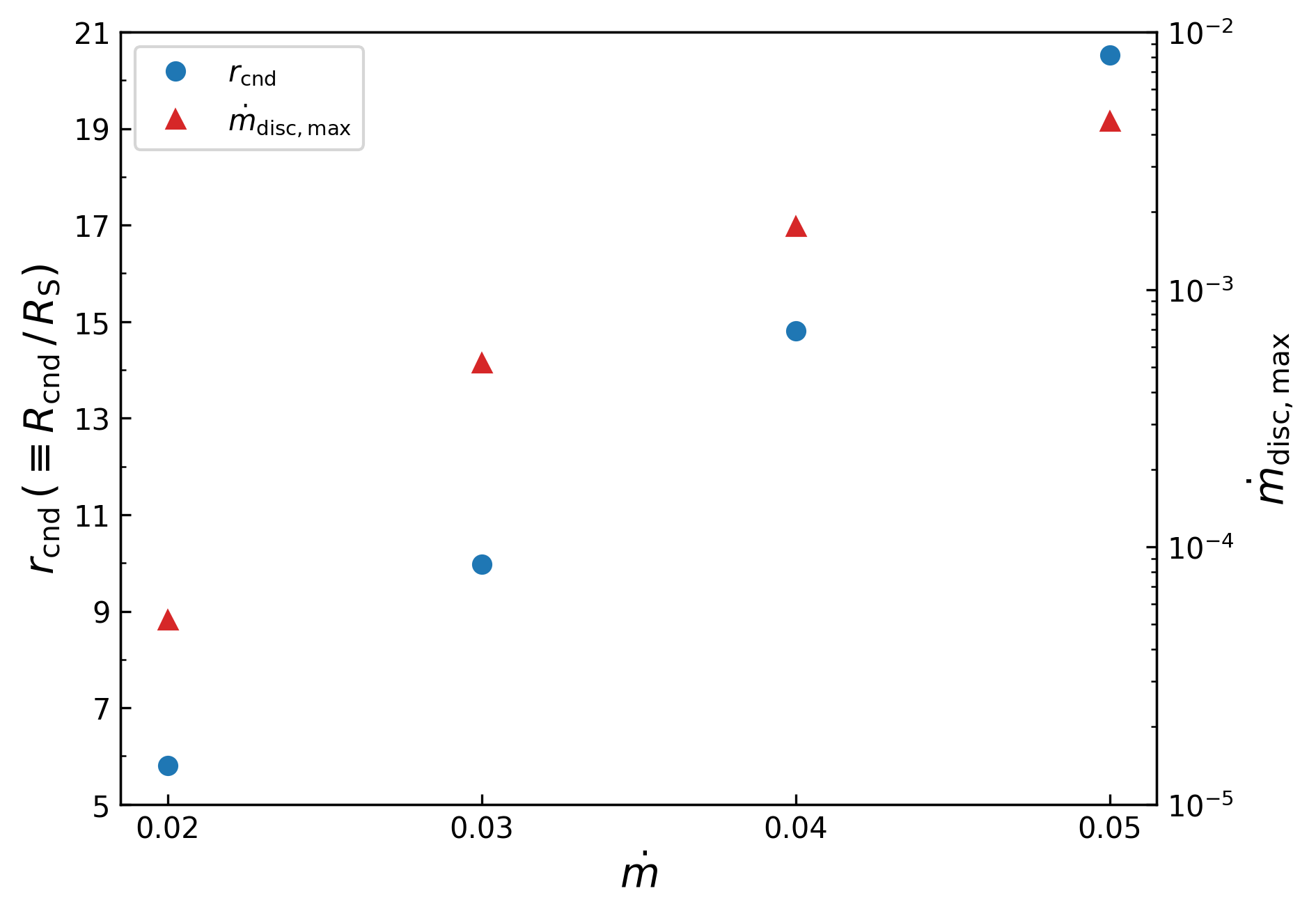}
    \caption{The condensation radius (blue dots) and the  accumulated disc accretion rate (red triangles) in dependence on the mass-supply rate $\dot{m}$.}
    \label{fig:r_cnd}
\end{figure}

The properties of the accretion flows from numerical computations are presented for a given viscous parameter, $\alpha = 0.4$, and the updated black hole mass of Cygnus X-1, $m = 21.2$  \citep[][]{miller-jones2021}. As the geometry of the accretion flows and the emission spectrum (except for a shift of disc component) are independent of the black hole mass \citep[e.g.][]{liu2007,qiao2018b} and  there is limited knowledge on the value of the viscous parameter, we concentrate on investigating the variation of the accretion geometry with respect to the mass-supply rate ($\dot{m}$). We perform calculations for a range of accretion rates and plot in the upper panel of Fig. \ref{fig:mdot} the radial distribution of the accretion rate in the corona, $\dot{m}_{\rm cor}(r)$, and in the disc, $\dot{m}_{\rm disc}(r)$, for a typical range of $\dot{m}$ with condensation. For clarity, the ratio of the disc accretion rate to the coronal accretion rate is also plotted in the lower panel. Both the accretion rate and the size of the disc increase with increasing $\dot m$, which is clearly shown in Fig. \ref{fig:r_cnd} by the critical condensation radius ($r_{\rm cnd}\equiv R_{\rm cnd}/R_{\rm S}$) and the accumulated disc accretion rate (i.e., the maximum disc accretion rate, $\dot{m}_{\rm disc,max}$) varying with the mass-supply rate. These dependences result from more efficient Coulomb coupling between the electrons and ions in the hot flows at higher accretion rates, which leads to more efficient radiative cooling and stronger condensation to the disc. It is also shown that the gas condensation is negligible for $\dot{m}=0.02$, implying that the hot flow remains as an ADAF at all distances for lower mass-supply rates.

\begin{figure}
    \centering
    \includegraphics[width=\linewidth]{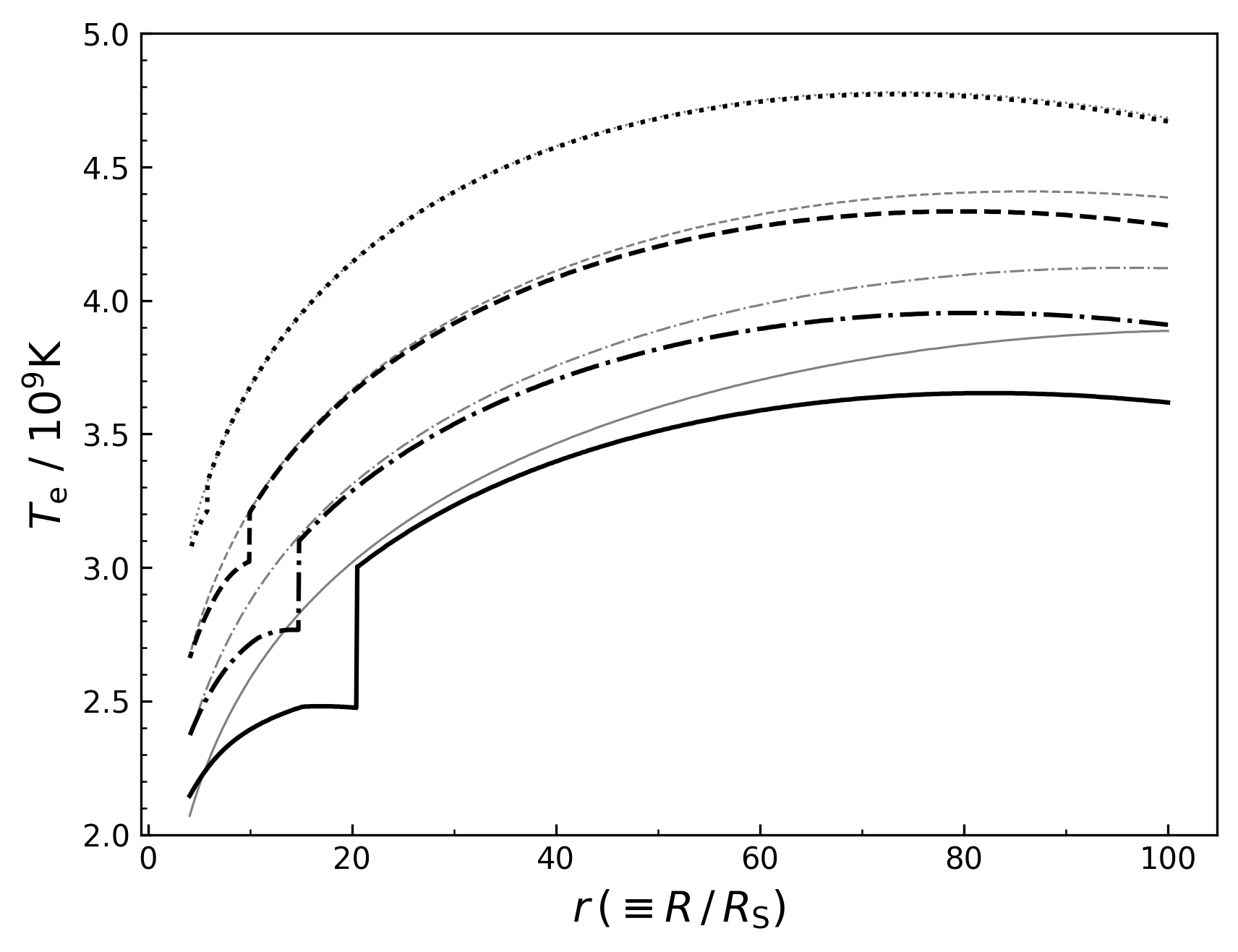}
    \caption{The radial distribution of electron temperature $T_{\rm e}$ in the hot accretion flows (thick lines) with mass-supply rate $\dot{m} = $ 0.05 (solid lines), 0.04 (dashed lines), 0.03 (dotted-dashed lines), 0.02 (dotted lines). For comparison the corresponding electron temperature in the pure ADAFs is shown by thin lines.}
    \label{fig:Te}
\end{figure}

In Fig. \ref{fig:Te} the radial distribution of the electron temperature in the corona for various $\dot{m}$ is plotted. For comparison, the corresponding electron temperature in a pure ADAF is also illustrated. The abrupt jump in temperature at the boundary between the corona and the outer hot flow, $r = r_{\rm cnd}$, is caused by extra cooling in the corona due to the Comptonization of soft photons from the underlying disc and the vertical conduction. Compared with the pure ADAFs (light, thin curves), the electron temperature in the coronae is slightly lower because of the existence of a disc. It approaches the electron temperature of the ADAFs when the condensation is negligibly weak and is essentially the same at very low accretion rates. Fig. \ref{fig:Te} shows a general trend, in both the corona and the ADAF, that the electron temperature decreases with increasing $\dot{m}$. This can be understood as the natural result of the increased radiative cooling efficiency due to the higher density in the accretion flow.

\begin{figure}
    \centering
    \includegraphics[width=\linewidth]{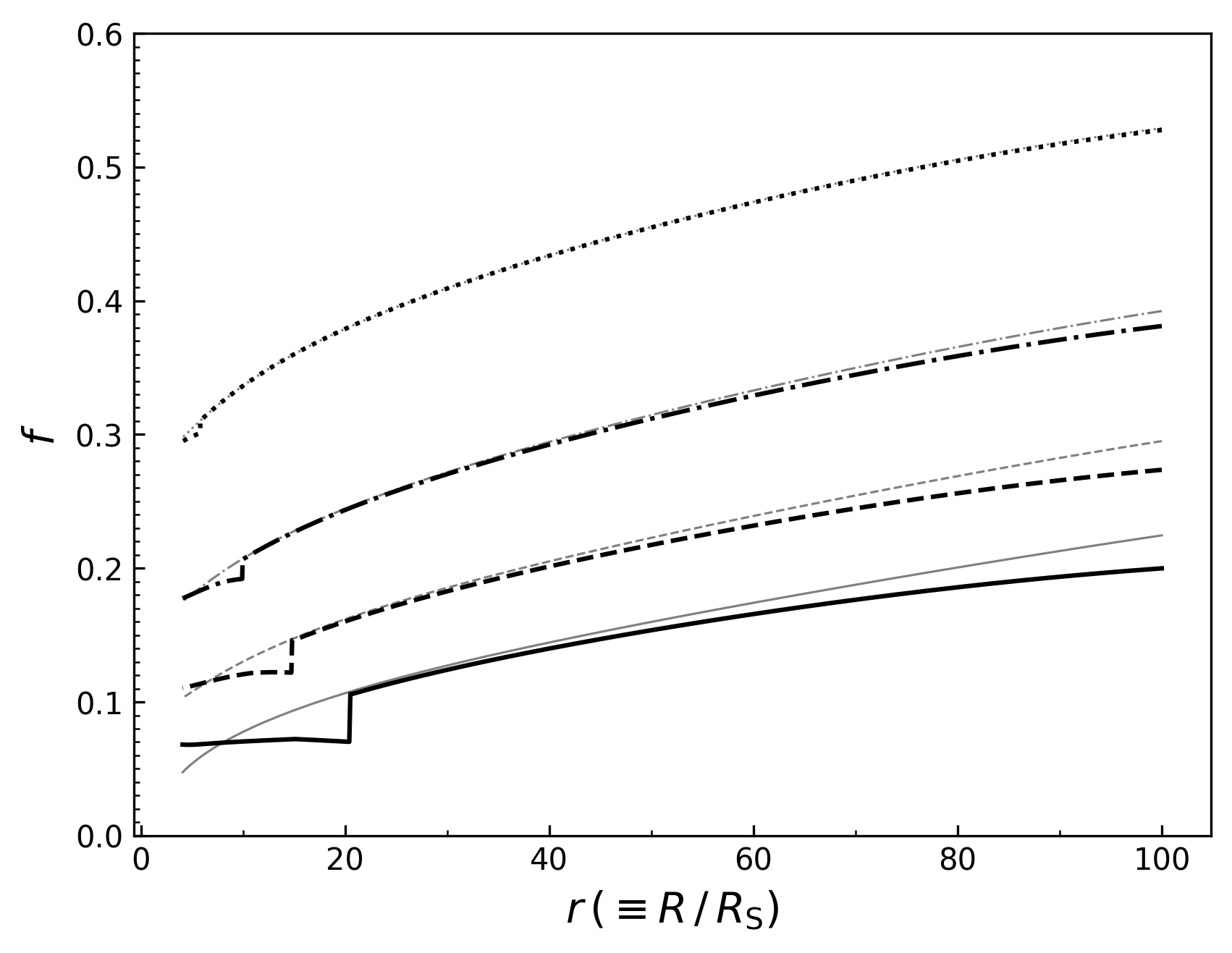}
    \caption{The radial distribution of the advection fraction $f$ in the hot accretion flows (thick lines) with mass-supply rate $\dot{m} = $ 0.05 (solid lines), 0.04 (dashed lines), 0.03 (dotted-dashed lines), 0.02 (dotted lines). The corresponding distribution of  $f$ in the pure ADAFs is shown by thin lines.}
    \label{fig:f}
\end{figure}

\begin{figure}
    \centering
    \includegraphics[width=\linewidth]{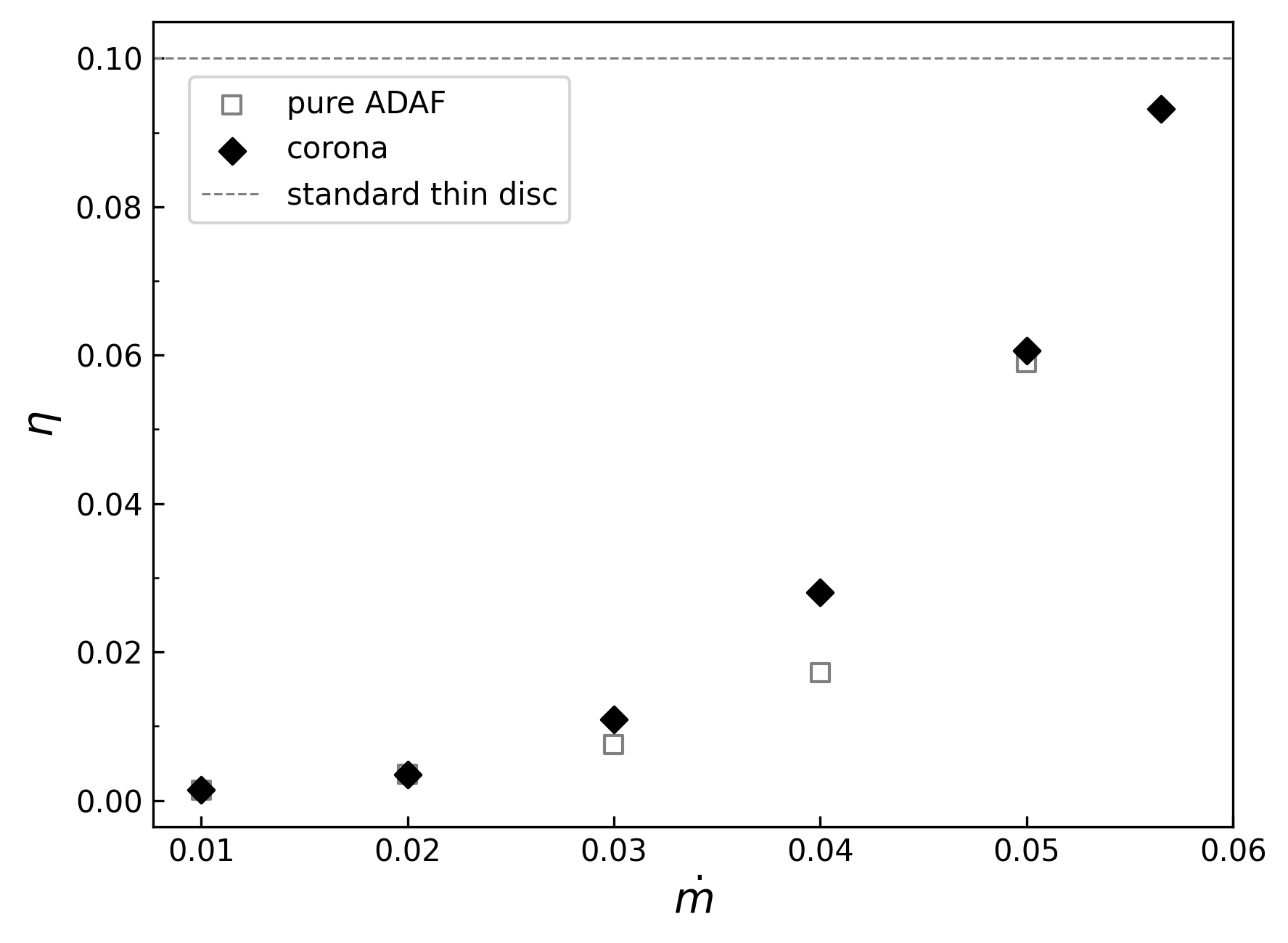}
    \caption{The radiative efficiency of the two-phase accretion flows (black diamonds) as a function of the mass-supply rate. At $\dot{m}=0.01$, no condensation occurs, and the accretion flow remains a pure ADAF at all distances. Unlike the standard thin disc, the radiative efficiency in the hot flows increases significantly with the mass-supply rate. For comparison the equivalent results for pure ADAFs are also shown as grey squares, except for $\dot{m}=0.057$ which is higher than the critical mass-supply rate of a pure ADAF (see Fig. \ref{fig:mdot_crit}).}
    \label{fig:eta}
\end{figure}

\begin{figure}
    \centering
    \includegraphics[width=\linewidth]{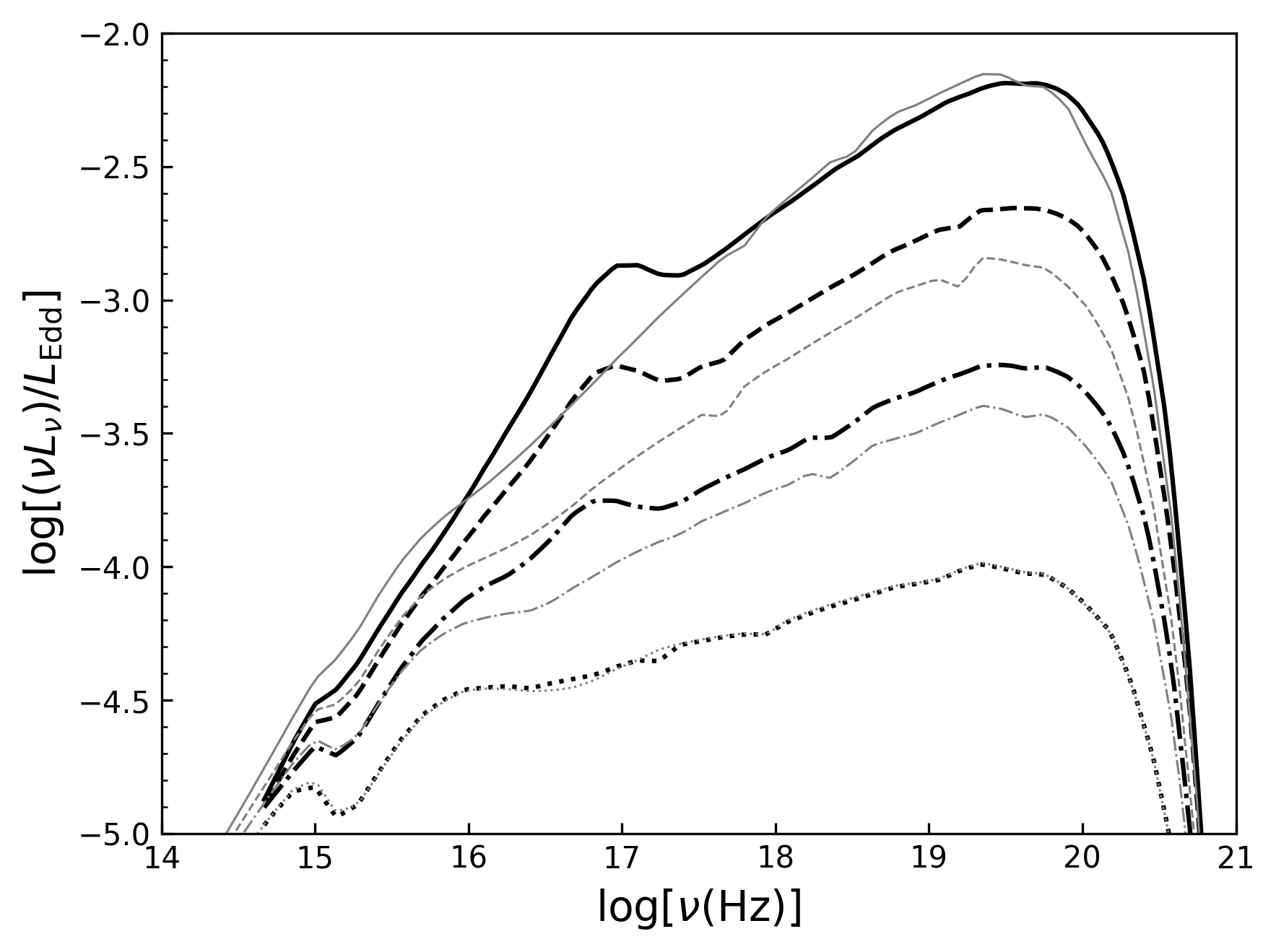}
    \caption{The emergent spectra of the two-phase accretion flows (thick lines) with mass-supply rate $\dot{m} = $ 0.05 (solid lines), 0.04 (dashed lines), 0.03 (dotted-dashed lines), 0.02 (dotted lines). For comparison the equivalent results for pure ADAFs are also shown as thin lines.}
    \label{fig:spectra}
\end{figure}

The self-consistent advection fraction of the viscous dissipation of energy, $f$, is plotted in Fig. \ref{fig:f}. With the gas flowing into a smaller volume, the Coulomb coupling becomes more efficient and a smaller fraction of the energy is advected, as shown by the radial distribution of the advection fraction in either the corona or the  ADAF. The slight difference in $f$ between the ADAF and corona is caused by the external Compton scatterings of the central and local disc photons, as well as by the vertical thermal conduction. Fig. \ref{fig:f} also shows that the advection fraction decreases with accretion rate ranging from 0.02 to 0.05. Note that while the value of $f$ measures the fraction of viscous heat to be advected in the form of entropy, the remaining fraction, $1-f$, does not represent the decreasing factor of the radiative efficiency on the basis of the standard disc. This is because the viscous dissipation rate also decreases with the increase of $f$, since a greater amount of gravitational work/energy is in the kinetic energy. As a consequence, the radiative efficiency decreases sharply with the increase of $f$ or the decrease of accretion rate, in contrast to the case of a constant value for a standard disc. We calculate $L$, the overall emission power from the disc and the corona, and plot the radiative efficiency, $\eta=L/\dot M c^2$, for different accretion rates in Fig. \ref{fig:eta}. As is shown, the radiative efficiency increases by more than one order of magnitude with the accretion rate changing from 0.01 to 0.05. The energy conversion efficiency is $\eta \approx 0.09$ when the accretion rate reaches the critical upper limit $\dot{m}_{\rm crit} \approx 0.057$ (see Fig. \ref{fig:mdot_crit} and related discussion in text), which is close to that of a standard thin disc. Such an increase in the radiative efficiency is a consequence of not only the decrease of advection fraction of the viscous heat (as is shown in Fig. \ref{fig:f}), but more importantly, the increase of the viscous dissipation rate at higher accretion rates. We have also plotted the equivalent results of pure ADAFs in Fig. \ref{fig:eta} (and later in Fig. \ref{fig:spectra} and Fig. \ref{fig:gamma_alpha0.4}), but we leave the comparison of the radiative properties of pure ADAFs and two-phase accretion flows for Section \ref{subsec: compare ADAF and corona}. 

\begin{figure}
    \centering
    \includegraphics[width=\linewidth]{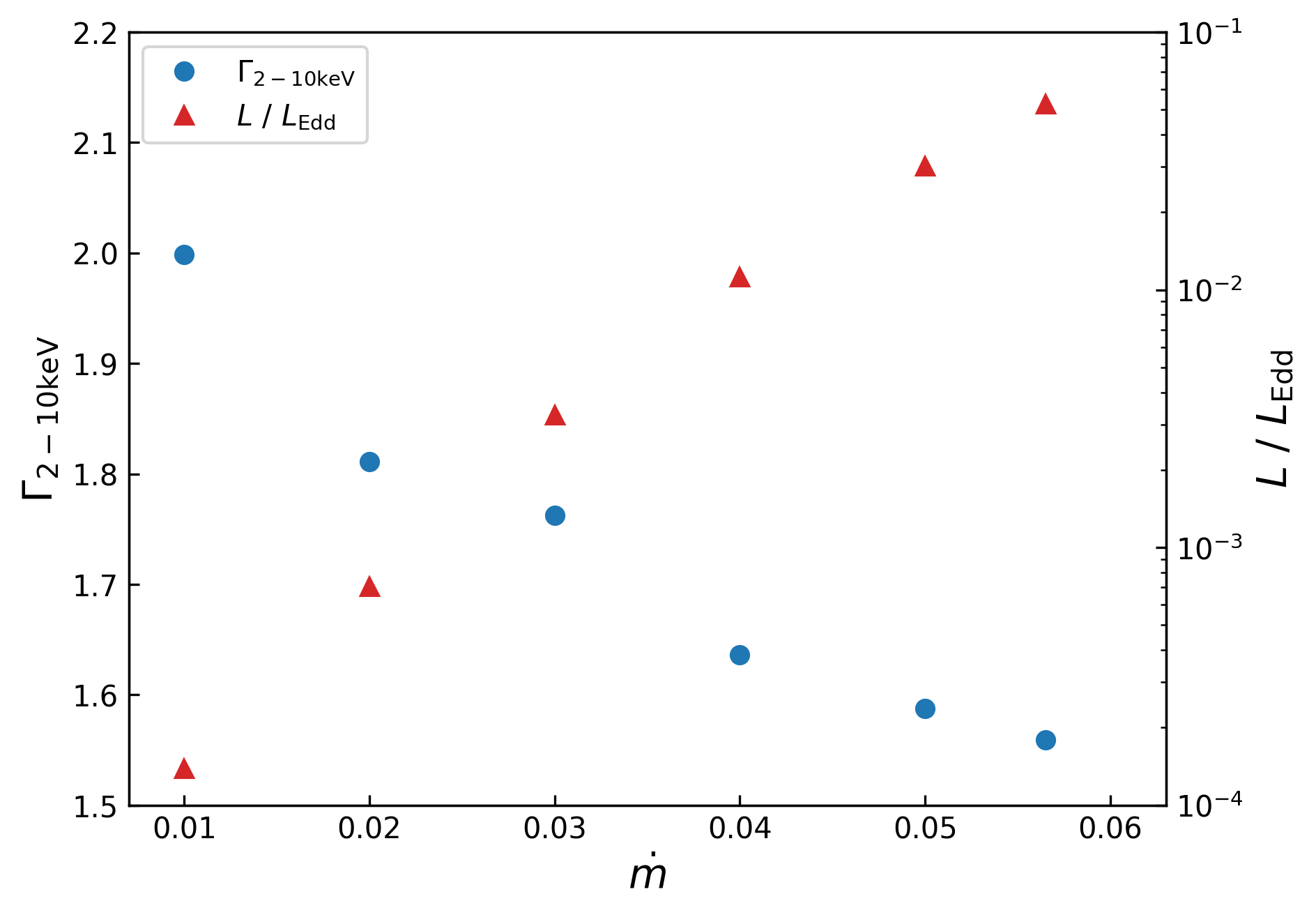}
    \caption{The Eddington ratio, $L / L_{\rm Edd}$ (red triangles), and the photon index between 2 and 10 keV, $\Gamma_{\rm 2-10 keV}$ (blue dots), of the two-phase accretion flows with different mass-supply rate. At $\dot{m}=0.01$ the accretion flow remains a pure ADAF at all distances.}
    \label{fig:luminosity}
\end{figure}

\begin{figure}
    \centering
    \includegraphics[width=\linewidth]{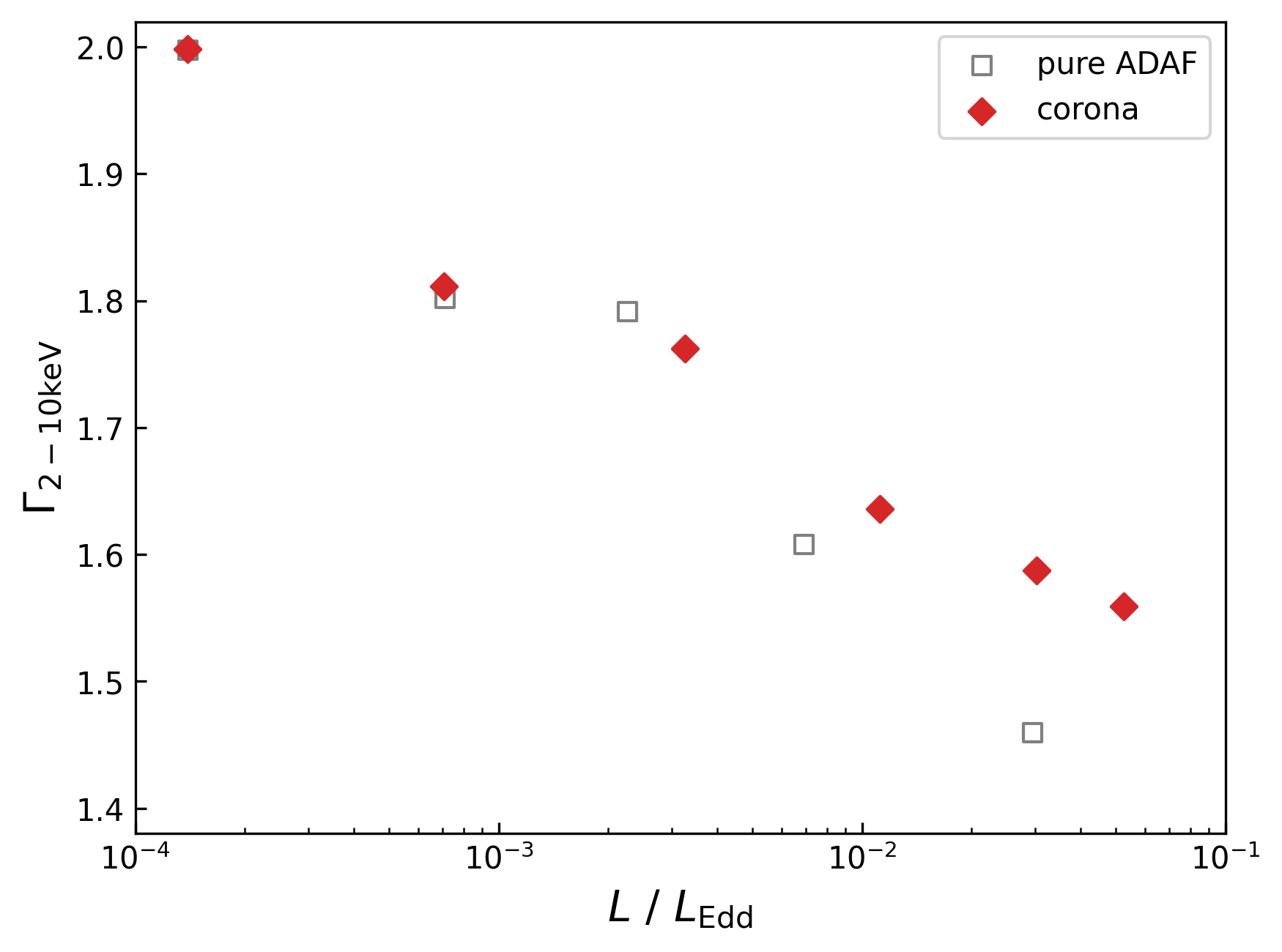}
    \caption{Dependence of the photon index between 2 and 10 keV on the Eddington ratio for the two-phase accretion flows (red diamonds). From left to right, the mass-supply rate is $\dot{m} = 0.01$ (no condensation), 0.02, 0.03, 0.04, 0.05, 0.057 ($\dot{m}_{\rm crit}$). For comparison the equivalent results for pure ADAFs are also shown as grey squares, except for $\dot{m}=0.057$ which is higher than the critical mass-supply rate of a pure ADAF (see Fig. \ref{fig:mdot_crit}).}
    \label{fig:gamma_alpha0.4}
\end{figure}

\begin{figure}
    \centering
    \includegraphics[width=\linewidth]{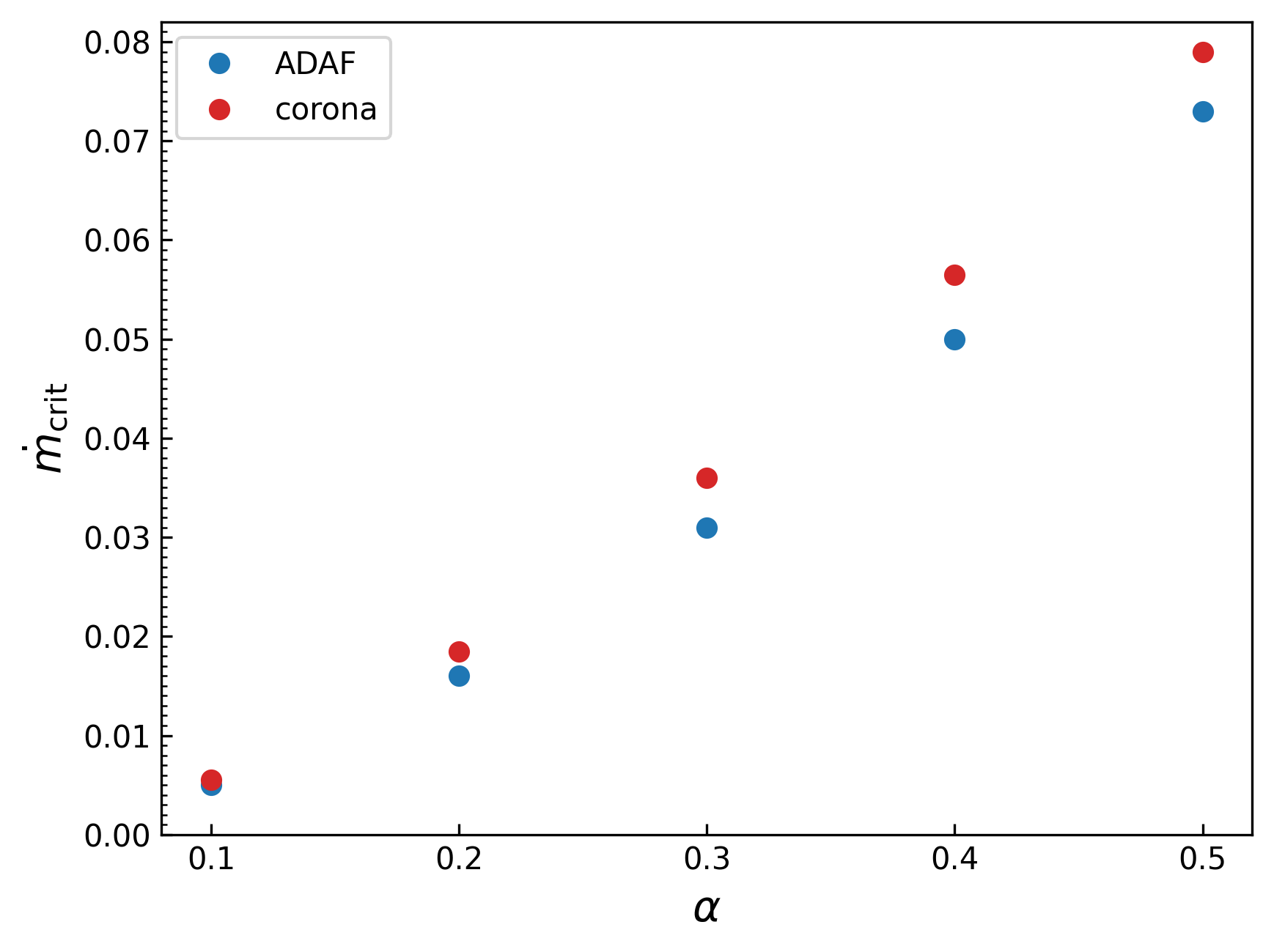}
    \caption{The dependence of the critical mass-supply  rate $\dot{m}_{\rm crit}$ on the viscosity parameter $\alpha$ for the hot accretion flows with condensation (red dots) and for the pure ADAFs (blue dots).}
    \label{fig:mdot_crit}
\end{figure}

In Fig. \ref{fig:spectra} we plot the emergent spectra of the two-phase accretion flows with $\dot{m}$ ranging from 0.02 to 0.05. With increasing $\dot{m}$, a clear multi-color black body bump occurs, contributed by the condensation-fed disc. In addition, the spectrum hardens and the luminosity increases substantially with the increase of mass-supply rate, as shown in Fig. \ref{fig:luminosity} illustrating the Eddington ratio $L/L_{\rm Edd}$ ($L_{\rm Edd} = 1.26 \times 10^{38} m$ is the Eddington luminosity) and the photon index between 2 and 10 keV ($\Gamma_{\rm 2-10 keV}$) as functions of the accretion rate. Fig. \ref{fig:luminosity} demonstrates that the Eddington ratio increases steeply with the accretion rate, in contrast to $L/L_{\rm Edd}=\dot{m}$ in the standard disc. The overall spectra are typical hard state spectra (see Fig. \ref{fig:spectra}), with the hard X-ray photon index ranging from 2.00 to 1.59 for $\dot{m}=0.01-0.05$ (see Fig. \ref{fig:luminosity}), with a minimum of 1.56 when the mass-supply rate reaches the upper limit, $\dot{m}=\dot{m}_{\rm crit} \approx 0.057$. Such X-ray spectra differ from our previous results \citep[e.g.][]{qiao2013a, qiao2018b} where the spectra never approached such a hardness from an accretion flow in a disc-corona configuration. The difference is a direct result of the improvement in the calculation of the energy density of the soft photons from the disc. The previous over-estimation led to a greater condensation and a generally softer emergent spectra at a given accretion rate. With the current modification, the predicted spectrum continuously hardens with increasing Eddington ratio, and remains hard at a sufficiently higher Eddington ratio. For example, $\Gamma_{\rm 2-10 keV}\approx 1.59$ at $L/ L_{\rm Edd}=0.03$ (Fig. \ref{fig:gamma_alpha0.4}), which is in better agreement with observations (see Section \ref{subsec: gamma-Eddington ratio}).

For the purpose of comparison with observations, the variation of the photon index $\Gamma_{\rm 2-10 keV}$ with the Eddington ratio $L/ L_{\rm Edd}$ is shown in Fig. \ref{fig:gamma_alpha0.4}. As can be seen, the model predicts typical hard state spectra for a reasonable range of the Eddington ratio. Combining the spectral feature and the thin disc size shown in Fig. \ref{fig:r_cnd}, our model allows for the presence of a broad Fe K$\alpha$ line in the hard state for an appropriate range of Eddington ratio as is discussed in Section \ref{subsec: inner disc}.

When the mass-supply rate exceeds a critical value, $\dot{m} > \dot{m}_{\rm crit}$, a two-temperature hot flow solution no longer exists because the Coulomb collisions are sufficiently efficient that the viscous heating of the ions is balanced by the radiative cooling of the electrons. The critical rate of the hot accretion flow considered here is essentially similar to that of the ADAF. The latter has been discussed in the literature \citep[e.g.][]{narayan1995b, narayan1998, mahadevan1997, Li2023} and was demonstrated to be dependent on the viscous parameter. We calculate the critical mass-supply rate of the hot accretion flow for a number of viscosity parameters, as well as that of the ADAF for comparison. As is shown in Fig. \ref{fig:mdot_crit}, the upper limit to the mass-supply rate for the hot flow to exist is slightly higher than that of the ADAF for the same viscous parameter. Such a result is caused by the gas condensation in the inner region, which diverts a small fraction of the hot accretion flow to the disc. Specifically, while the Coulomb collisions become efficient to avert the ADAF at the innermost region, the condensation reduces the density, making it possible for the hot corona to exist, which connects to an outer ADAF where the corresponding critical accretion rate can be slightly higher. 
 
It is found that the critical accretion rate, either for the ADAF or for the hot flow with condensation, is sensitive to the viscosity. A large viscous parameter implies rapid inflow and hence low density for a fixed accretion rate. The low density weakens the effect of Coulomb collisions, and reduces the radiation rate. Both effects enhance the effect of advection. Therefore, in the hot accretion flows or the pure ADAFs, a larger $\alpha$ allows a higher $\dot{m}_{\rm crit}$. As is shown in Fig. \ref{fig:mdot_crit}, the critical mass-supply rate for the hot accretion flow can be as high as $\sim 0.08$ when $\alpha=0.5$. The increased difference in the critical accretion rates between the hot accretion flows and the ADAFs with increasing $\alpha$ is caused by the enhanced condensation at high accretion rates. 

\begin{figure}
    \centering
    \includegraphics[width=\linewidth]{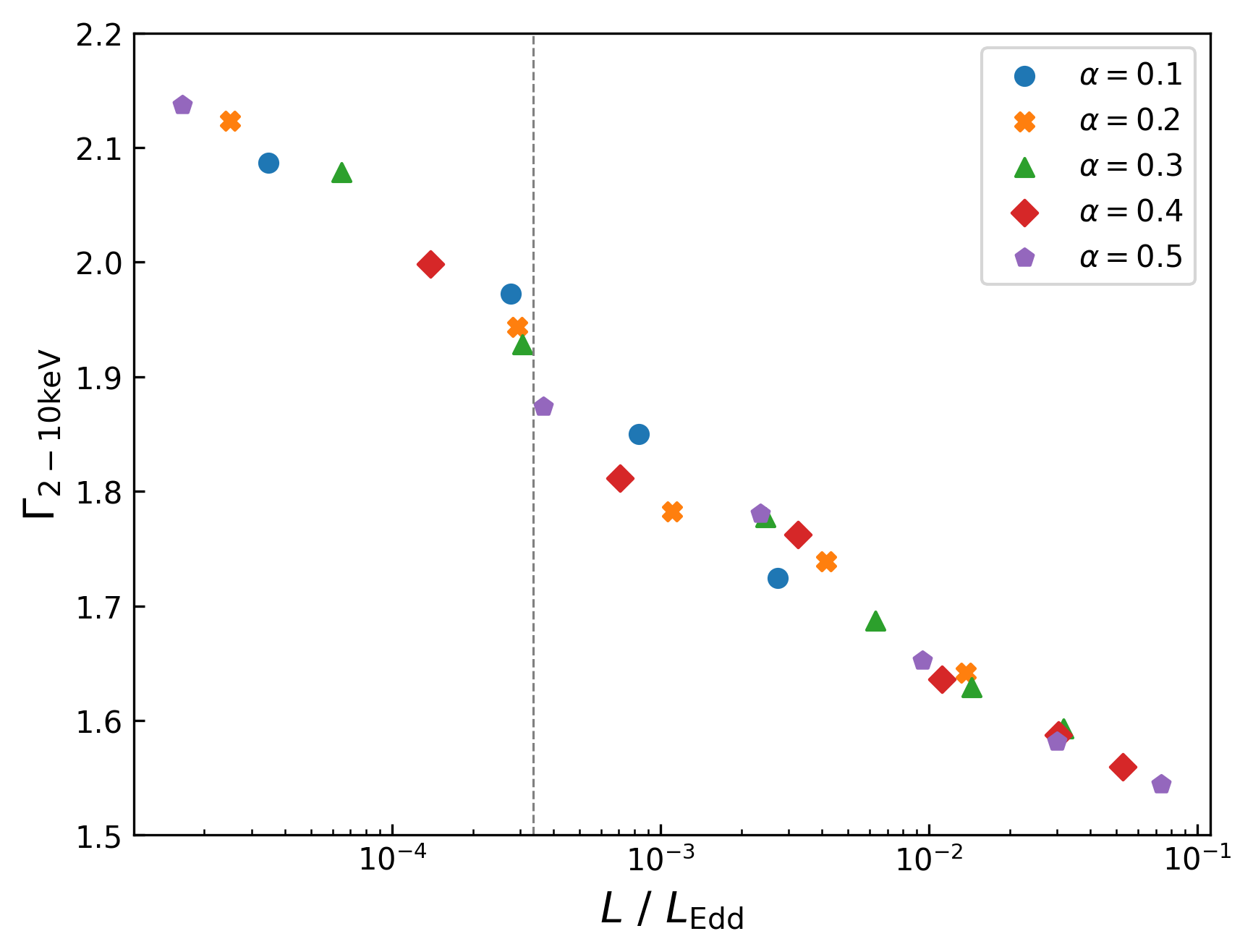}
    \caption{Dependence of the photon index between 2 and 10 keV on the Eddington ratio for viscosity parameter $\alpha = 0.1$ (blue dots), 0.2 (orange crosses), 0.3 (green triangles), 0.4 (red diamonds), 0.5 (purple pentagons). The mass-supply rate is, from left to right, for $\alpha = 0.1$, $\dot{m} = 0.001$, 0.003, 0.004, 0.006 ($\dot{m}_{\rm crit}$), for $\alpha = 0.2$, $\dot{m} = 0.002$, 0.006, 0.01, 0.014, 0.019 ($\dot{m}_{\rm crit}$), for $\alpha = 0.3$, $\dot{m} = 0.005$, 0.01, 0.02, 0.025, 0.03, 0.036 ($\dot{m}_{\rm crit}$), for $\alpha = 0.4$, $\dot{m} = 0.01$, 0.02, 0.03, 0.04, 0.05, 0.057 ($\dot{m}_{\rm crit}$), and for $\alpha = 0.5$, $\dot{m} = 0.005$, 0.02, 0.036, 0.05, 0.065, 0.079 ($\dot{m}_{\rm crit}$). The vertical dashed line marks the lowest Eddington ratio for the coronal condensation to take place.}
    \label{fig:gamma}
\end{figure}

While the viscosity parameter affects the upper limit of the Eddington ratio of a hard spectral state (via critical supply rate), it is interesting to note that there is a degeneracy in $\alpha$ and $\dot m$ in producing a spectrum. As is shown in Fig. \ref{fig:gamma}, a given spectrum, represented by its photon index and Eddington ratio, can be produced by different combinations of $\alpha$ and $\dot m$. For example, a spectrum with $\Gamma_{\rm 2-10 keV}\sim 1.6$ and $L / L_{\rm Edd} \sim 0.03$ (see Fig. \ref{fig:gamma}) can be emitted by an accretion flow with $\alpha=0.5, \dot m=0.065$, or $\alpha=0.4, \dot m=0.05$, or $\alpha=0.3, \dot m=0.036$. Fig. \ref{fig:gamma} shows that the relation between the spectral shape (photon index) and the Eddington ratio falls along the same curve for different viscosity parameters, thus, reducing the freedom of choice for parameters in the fitting of an observed spectrum. 

For a hot gas supply rate greater than the critical accretion rate, condensation caused by the energy balance is insufficient to steadily divert enough gas such that the accretion rate in the corona decreases below $\dot m_{\rm crit}$, and thus, no steady hot accretion flows can be found. In LMXBs, such a problem does not exist as the hot gas supply rate cannot exceed the maximum evaporation rate, while in the wind-fed systems there is no apparent upper limit for the wind supply rate. However, one would anticipate that an over-critical hot gas fed by a stellar wind could partially collapse into a disc as a result of cooling in disturbances at large distances, which coexists with a residual, steady hot flow. It is also possible that the supplied gas totally collapses into a thin disc at the outer boundary, and a corona forms by evaporation. In these cases, a spectral transition to a soft state, dominated by the disc emission, would be possible. We leave the investigation of the soft state for our future work. 

\section{Comparison with observations}
\label{sec: comparison with observations}

\subsection{Spectrum of Cygnus X-1 in the low/hard state}

\begin{figure}
    \centering
    \includegraphics[width=\linewidth]{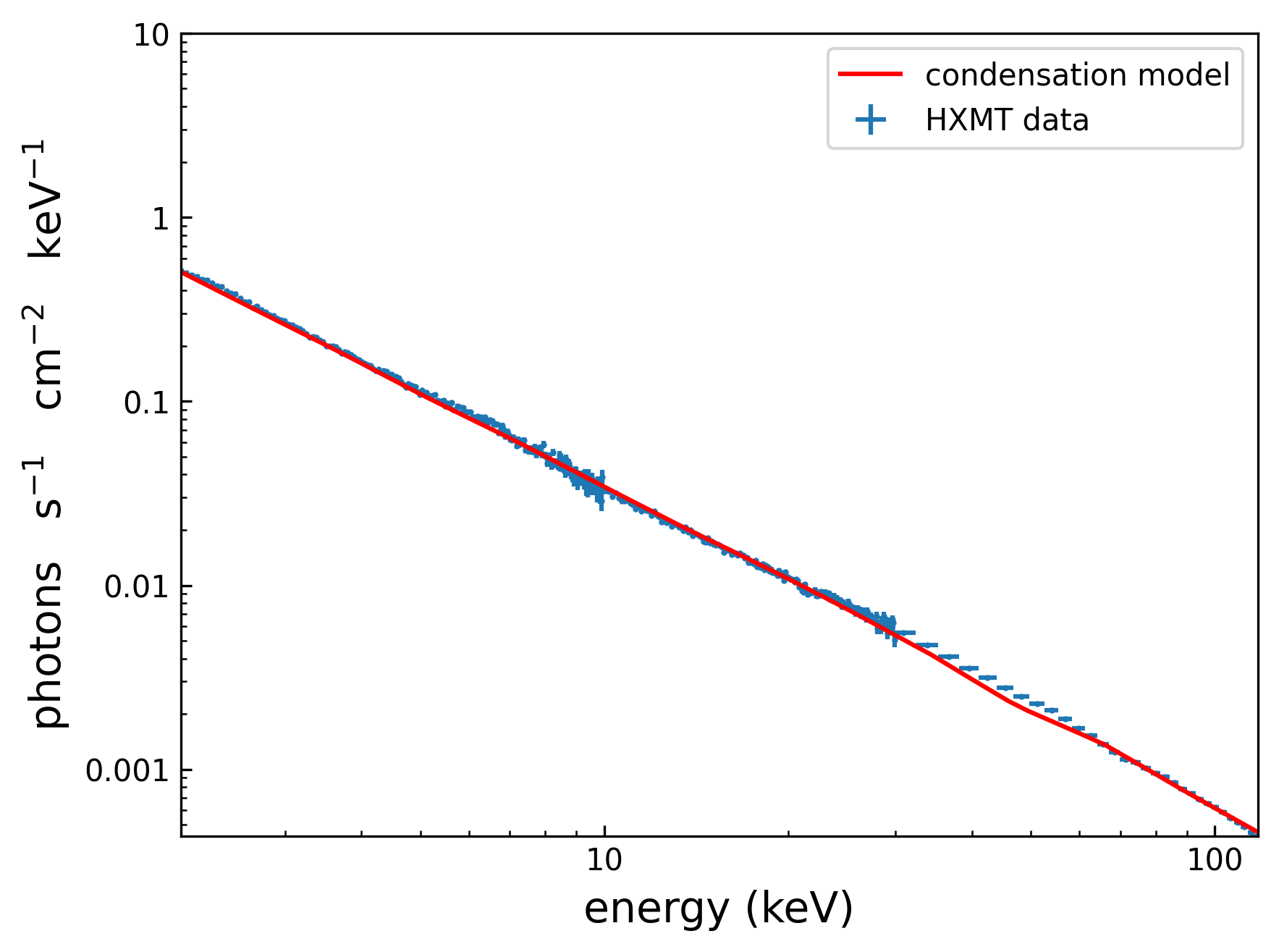}
    \caption{The theoretical (red solid line) and observed (blue points with errors) energy spectra of Cygnus X-1 in the hard spectral state. The theoretical results are calculated with mass-supply rate $\dot{m} = 0.026$ and viscosity parameter $\alpha = 0.3$. The observed energy spectrum is derived by \citet{feng2022} using the HXMT data.}
    \label{fig:cyg X-1 spectrum}
\end{figure}

Cygnus X-1 is a typical HMXB where the black hole accretes via the stellar wind of its supergiant companion \citep[e.g.][]{Lamers1976}. \citet{feng2022} obtained the broadband energy spectrum of Cygnus X-1 in the hard spectral state using the HXMT data. As an example, we apply the condensation model to Cygnus X-1 and compare the theoretical spectrum to the HXMT observations. We take $m = 21.2$, the latest estimate of the black hole mass of Cygnus X-1 \citep{miller-jones2021}, and $\alpha = 0.3$, as suggested by \citet{king2007}. For a set of models corresponding to a range of accretion rates we derive the structure of the two-phase accretion flow and calculate their theoretical broad band energy spectrum. Taking into account the distance $d$ and the orbital inclination $i$ of Cygnus X-1, which are given as $d \approx 2.22~\rm{kpc}$ and $i \approx 27.51^{\circ}$ by \citet{miller-jones2021}, the theoretical spectrum is compared with the observed one. In Fig. \ref{fig:cyg X-1 spectrum} we overlay the theoretical spectrum corresponding to $\dot m=0.026$ and the one observed by \citet{feng2022}. It can be seen that the corona condensation model offers a promising explanation to this hard state spectrum of Cygnus X-1.

\subsection{Inner disc in the low/hard state}
\label{subsec: inner disc}

It is commonly thought that the standard disc is truncated in the low/hard state and the inner region is described by an ADAF to produce the hard spectrum \citep[e.g.][]{esinetal1997}. Nevertheless, for a number of BHXBs evidence has been found that an optically thick accretion disc extends close to the ISCO in the low/hard state. The evidence for its existence is inferred by modeling either the thermal component of the X-ray spectrum or the relativistic blurring of the Fe K$\alpha$ line \citep[e.g.][]{miller2006, miller2015, reis2010, parker2015, poutanen2018, garcia2018, kara2019,buisson2019, ren2022, dong2022}. This interpretation challenges the conventional accretion theory since the continuum spectrum of a standard thin disc extending to the ISCO is too soft while a truncated disc fails to produce the reflected broad Fe K$\alpha$ line.
 
The coupled disc-corona scenario provides a promising explanation. As we have demonstrated in Section \ref{sec:Numerical results}, the steady gas condensation, as a consequence of disc-corona interaction, supports a weak inner disc around the ISCO while the accretion flow is dominated by the hot corona. Such an accretion geometry exists for an Eddington ratio ranging from  $\sim 10^{-3}$ to $\sim 0.1$, with the upper limit depending on the value of the viscosity parameter $\alpha$ (see Fig. \ref{fig:gamma}). We remark that the condensation model is applicable to both the HMXBs and the LMXBs in the low/hard state, and therefore the existence of a condensation-fed inner disc is possible in both types of systems.

\subsection{Correlation between the hard X-ray photon index and the Eddington ratio}
\label{subsec: gamma-Eddington ratio}

The phenomenological properties of the observed hard X-ray spectra of BHXBs can be studied in terms of the photon index, $\Gamma$, and the Eddington ratio, $L / L_{\rm Edd}$. It has been found that a V-shaped correlation exists between $\Gamma$ and $L / L_{\rm Edd}$ \citep[e.g.][]{yuanf2007, wu2008, cao2014, yang2015, jana2022}, which is speculated to be triggered by a transition from the hot-flow-dominated hard state to the disc-dominated soft state. The turning point occurs at a different Eddington ratio in individual sources, with a smaller photon index at a higher Eddington ratio. For example, turning occurs with $\Gamma \sim 1.4$ at a flux of $10^{-9} \ {\rm ergs\ cm^{-2}\ s^{-1}}$ in the 3-9 keV band for GX 339-4 and with $\Gamma\sim 1.8$ at a flux of $10^{-9.6}\ {\rm ergs\ cm^{-2}\ s^{-1}}$ for H 1743-322 \citep[see Figure 5. of][]{cao2014}. This is qualitatively consistent with our disc-corona interaction model, which, as is shown in Fig. \ref{fig:gamma}, predicts a general relation in the hard state where $\Gamma$ decreases with increasing Eddington ratio and ends (turns) at a critical Eddington ratio depending on the viscous parameter. The positive correlation at higher Eddington ratio is not plotted in the figure, as the ADAF does not strictly exist.

\begin{figure}
    \centering
    \includegraphics[width=\linewidth]{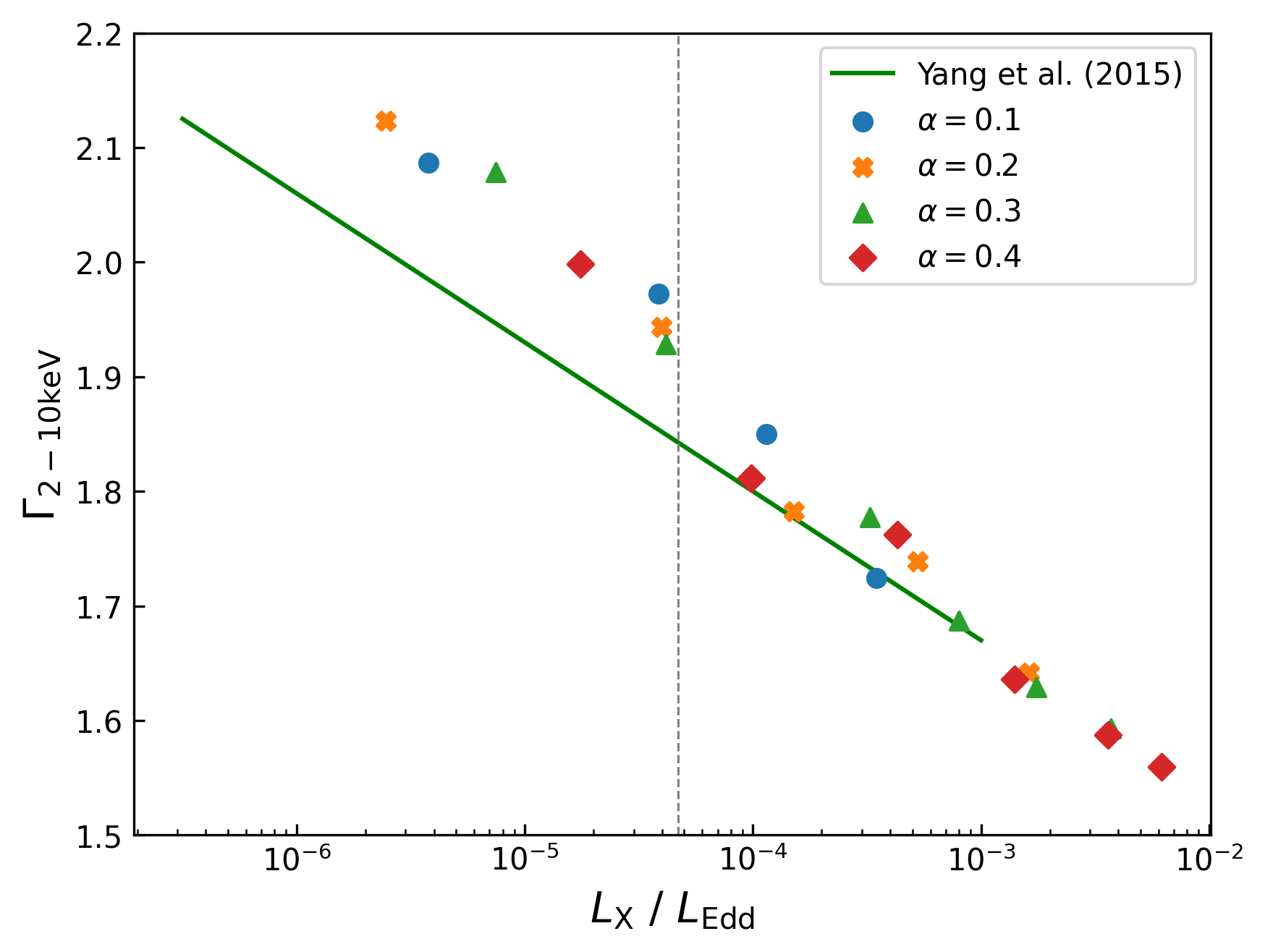}
    \caption{The correlation between the hard X-ray photon index, $\Gamma_{\rm 2-10 keV}$, and the 2-10 keV Eddington ratio, $L_{\rm X}/L_{\rm Edd}$. The green solid line, $\Gamma_{\rm 2-10 keV} = (-0.13\pm0.01){\rm log_{10}}(L_{\rm X}/L_{\rm Edd}) + (1.28\pm0.02)$, is the hard state branch ($10^{-6.5} \lesssim L_{\rm X} / L_{\rm Edd} \lesssim 10^{-3}$) of the linear fit obtained by \citet{yang2015} with the observational data of 13 BHXBs (uncertainties are at $1\sigma$ level). Scatter data are the numerical results of the corona condensation model, with $\alpha = 0.1\ $(blue dots), 0.2 (orange crosses), 0.3 (green triangles), 0.4 (red diamonds). The mass-supply rate is, from left to right, for $\alpha = 0.1$, $\dot{m} = 0.001$, 0.003, 0.004, 0.006 ($\dot{m}_{\rm crit}$), for $\alpha = 0.2$, $\dot{m} = 0.002$, 0.006, 0.01, 0.014, 0.019 ($\dot{m}_{\rm crit}$), for $\alpha = 0.3$, $\dot{m} = 0.005$, 0.01, 0.02, 0.025, 0.03, 0.036 ($\dot{m}_{\rm crit}$), and for $\alpha = 0.4$, $\dot{m} = 0.01$, 0.02, 0.03, 0.04, 0.05, 0.057 ($\dot{m}_{\rm crit}$). The vertical dashed line marks the lowest Eddington ratio for the coronal condensation to take place.}
    \label{fig:gamma_L_Yang}
\end{figure}

\citet{yang2015} analysed the spectral properties of 13 BHXBs and performed ${\rm least}-\chi^2$ linear fitting between the photon index and the X-ray luminosity in the 2-10 keV band ($L_{\rm X}$) in three branches characterised by different ranges of Eddington ratio \citep[see Figure 1. in][]{yang2015}, which allows us to compare our theoretical prediction with their fit of the hard state branch. The parameters, $m=21.2$, $\beta = 0.95$, $a=0.15$, are fixed in the computation. We have plotted the numerical results with $\alpha$ ranging from 0.1 to 0.4 \citep{king2007} and a series of $\dot{m}$ up to the corresponding critical values. From Fig. \ref{fig:gamma_L_Yang} it is clear that the disc-corona model predicts a similar (but slightly less steep) $\Gamma_{\rm 2-10 keV}-L_{\rm X} / L_{\rm Edd}$ correlation as suggested by observations, though there exists a large scatter among different sources. If we adopt the turning point at  $L_{\rm X} \sim 10^{-3} L_{\rm Edd}$, $\alpha \sim 0.2$ is favored for the hot corona. 

\section{Summary and Discussion}
\label{sec:summary}

In this work, we improve the numerical calculation procedure of the corona condensation model by using an iterative method to self-consistently calculate the advection fraction of the viscously dissipated energy at different radii of the hot accretion flow, along with other physical quantities which determine the structure of the two-phase accretion flow. In addition, we modified the energy density of the soft photons for Compton scattering, which plays a key role in the cooling of the corona and the emission of the hard X-ray spectrum. Specifically, we abandon the approximation of $\dot{m}_{\rm disc}(r) \approx \dot{m}_{\rm disc,max}$ in calculating the local disc emission, which over-estimated the external Compton cooling of the corona and consequently led to a higher condensation rate and much softer emergent spectrum. On the other hand, we include the contribution of the central disc emission to the soft photons for Compton scattering and adopt a lamp-post coronal illumination to the disc. 

Assuming an ADAF-like hot flow at $\sim 100 R_{\rm S}$, supplied either by the accretion of hot gas from the stellar wind or the complete evaporation of a disc as predicted by the disc evaporation model, we calculate the possible condensation as a consequence of disc-corona interaction and determine the accretion geometry at equilibrium. We find that in the low/hard state the structure and the radiation spectrum of the hot flows are generally similar to those of the ADAFs, except that, at an Eddington ratio $\gtrsim 10^{-3}$, there exists an inner thin disc fed by the condensation, which contributes a weak multi-color black body component in the overall spectrum. The modified corona condensation model produces almost exclusively hard emergent spectra, which provides an explanation of the observed hard spectral state for both the HMXBs and the LMXBs. As an example, we apply the model to Cygnus X-1 and demonstrate that the theoretically derived emergent spectrum shows a resemblance to the observed one. It is also found that the predicted relation between the photon index and the Eddington ratio is in good agreement with the observed evolution of individual sources and the statistical correlation for a sample of sources. One of the most important consequences of coronal condensation is that a weak inner disc can survive in the low/hard state, which provides a promising explanation for the puzzling existence of the relativistically blurred iron lines occurring in the low/hard state, complementing the conventional accretion scenario of the low/hard state and high/soft state and their transitions. 

In our future work, we intend to generalize the model by adopting a boundary condition with both hot and cold gas supplies to describe the accretion in a variety of astrophysical environments, leaving the wind supply and the RLOF supply as two extreme cases.

Other issues which need further clarification are discussed as follows.

\subsection{The approximation of coronal irradiation}
\label{subsec: lamp-post}

As was mentioned in Section \ref{subsec: the corona}, when calculating the coronal irradiation, we simplify the corona as a point source located at a height of $H_{\rm s} = 10R_{\rm S}$ above the black hole. The lamp-post simplification of our extended corona is reasonable because the coronal radiation is centrally concentrated, i.e., most of the hard X-ray radiation is emitted from within $\sim 10R_{\rm S}$ from the black hole \citep[e.g.][]{liu2017,reis2013,fabian2015}. The choice of coronal height, $H_{\rm s} = 10R_{\rm S}$, has been justified in our earlier works by theoretical approach \citep[][]{liu2002} and by comparison of model prediction with observation \citep[][]{qiao2017}. Although these justifications are made for AGNs, we continue to use this lamp-post approximation and the coronal height in the present work due to the analogy between AGNs and BHXBs. Moreover, we found that the model predictions are not sensitive to the lamp-post coronal height. For instance, with other parameters fixed to typical values ($m=21.2$, $\alpha = 0.4$, $\dot{m} = 0.03$), changing the height from $H_{\rm s} = 10 R_{\rm S}$ to $H_{\rm s} = 5 R_{\rm S}$ would only increase the maximum disc accretion rate ($\dot{m}_{\rm disc,max}$) by $\sim 1.5\%$, and change in the condensation radius ($R_{\rm cnd}$) is negligible.

We have also neglected the light bending effect when calculating the coronal irradiation. Light bending tends to increase the coronal irradiation and hence disc accretion rate in the innermost region, while it is less effective in the regions more distant from the black hole. Therefore, although light bending is expected to strengthen the thin disc near the ISCO, it has limited influence on the entire accretion flow and thus does not change our main conclusions.

\subsection{The uniqueness of the solution}
\label{subsec: solution uniqueness}

It has been well known that the accretion flow, either in the form of ADAF or standard disc, is uniquely determined for given $\alpha$, $\beta$, $a$, $m$, and $\dot{m}$. When the condensation is taken into account, a question arises as whether the condensation rate is unique to assure a unique disc-corona configuration. The answer is affirmed by our complete set of equations describing the corona and disc, supplemented by the condensation via the transition layer. This can also be understood with our iterative calculation method. In the computing procedure, a central corona luminosity and a disc luminosity are initially assumed for calculating the illumination and external Compton scatterings respectively. Then, from the outer boundary inwards, the corona and disc structure is calculated for each radius.  Similar to the standard disc, the solution for the local disc is solely determined by taking a test accretion rate in the disc, and then the solution for the corona can also be determined. Thus, the condensation rate can be derived as described in Section \ref{s:trans} and hence the disc accretion rate. This accretion rate is compared with the test value and iterative calculation continues until the presumed accretion rate equals to the derived one. Repeating above calculations at each radius, we derive the total corona  and disc luminosities, and compare them with the presumed values. If they are not consistent, we assume a new set of luminosities for the corona and disc, and repeat all above computations until a true solution is obtained. During the iterative calculations, three quantities, namely the disc accretion rate and the luminosities of the corona and disc, are determined by comparing the presumed and derived values. Since the derived values are all positively correlated with the corresponding presumed values (e.g. higher presumed disc accretion rate increases the external Compton cooling and hence enhances condensation), there can only be one, if any, self-consistent solution.

\subsection{Comparison of the radiative properties of pure ADAF and two-phase accretion flow}
\label{subsec: compare ADAF and corona}

In Fig. \ref{fig:eta}, Fig. \ref{fig:spectra} and Fig. \ref{fig:gamma_alpha0.4} we have plotted the radiative efficiency, emergent spectra and the $\Gamma_{\rm 2-10keV}-L/L_{\rm Edd}$ correlation of pure ADAFs with the same parameters as the two-phase accretion flows (except for $\dot{m} = 0.057$).

Fig. \ref{fig:eta} shows that the radiative efficiency of a pure ADAF is slightly lower than that of the two-phase accretion flow, which is particularly clear for a mass-supply rate of $\dot{m} \approx 0.04$. This can be understood because in the two-phase accretion flow a part of the relatively radiation-inefficient corona is condensed into a radiation-efficient thin disc. However, the two are essentially the same at either lower or higher accretion rates. The similarity at low accretion rates (e.g. $\dot{m} \lesssim 0.02$) is, of course, expected since the condensation of the corona is negligible. The similarity at higher accretion rates, on the other hand, is also natural because when the accretion rate is as high as $\dot{m} = 0.05$, the radiation in ADAF itself has become nearly as efficient as the thin disc and therefore the existence of a disc no longer makes a difference in the radiative efficiency.

However, although the radiative efficiencies of two-phase accretion flow and pure ADAF are similar at higher accretion rates, the spectral shapes differ. In the two-phase accretion flow, a thermal component is present in addition to the spectrum of the hot flow (Fig. \ref{fig:spectra}) and the coronal spectrum is also slightly softer than that of a pure ADAF with similar Eddington ratio (Fig. \ref{fig:gamma_alpha0.4}), as a consequence of the existence of the disc.

Moreover, as the critical mass-supply rate of the two-phase accretion flow is higher than that of a pure ADAF with the same $\alpha$ (Fig. \ref{fig:mdot_crit}), the condensation model allows higher maximum Eddington ratio in the hard state of BHXBs, which is shown in Fig. \ref{fig:gamma_alpha0.4}.  

\subsection{The effect of varying disc size on the Fe K$\alpha$ line profile}

Unlike a conventional thin disc, the condensation-fed disc has a varying outer radius ($R_{\rm cnd}$). Since the observed broad Fe K$\alpha$ line is the sum of the line emission from all radii of the disc, the varying $R_{\rm cnd}$ should have an impact on the line profile.

As is discussed in e.g. \citet{fabian2000}, at each radius in the rotating disc, the non-relativistic Doppler effect produces a symmetric double-peaked line profile, of which the blue peak is  enhanced and the red peak is weakened due to the beaming effect. The transverse Doppler effect and the gravitational redshift further move the line to a lower energy. As compared to the outer regions of the disc, the innermost part contributes to the broadest part of the line, as it has the highest orbital velocity, and for the same reason the enhancement to the blue peak of the iron line produced by the innermost part is also more effective.

Therefore, with the increase of the accretion rate during the outburst of a BHXB, the outer radius of the disc increasing from the ISCO to e.g. $\sim 20 R_{\rm S}$ (Fig. \ref{fig:r_cnd}) could cause a changing Fe K$\alpha$ line profile with initially sharper peaks being blurred as the system evolves. It is also likely that the blue peak of the line is much stronger than the red peak in the earlier phase, but less so in the later phase. Moreover, the width of the broad line is expected to stay the same because the inner radius of the disc is a constant.

\section*{Acknowledgements}

The authors are grateful to R. E. Taam for valuable discussions and improvements on the manuscript. Y. L. Wang thanks M. Z. Feng for providing the HXMT data. This work is supported by NSFC grants 12073037, 12203071, and 12333004.

%%%%%%%%%%%%%%%%%%%%%%%%%%%%%%%%%%%%%%%%%%%%%%%%%%
\section*{Data Availability}

The data and code used in this work will be shared on reasonable request to the corresponding authors.
 
%The inclusion of a Data Availability Statement is a requirement for articles %published in MNRAS. Data Availability Statements provide a standardised format for %readers to understand the availability of data underlying the research results %described in the article. The statement may refer to original data generated in the %course of the study or to third-party data analysed in the article. The statement %should describe and provide means of access, where possible, by linking to the data %or providing the required accession numbers for the relevant databases or DOIs.



\bibliographystyle{mnras}
\bibliography{accretion} % if your bibtex file is called example.bib

%%%%%%%%%%%%%%%%%%%%%%%%%%%%%%%%%%%%%%%%%%%%%%%%%%

%%%%%%%%%%%%%%%%% APPENDICES %%%%%%%%%%%%%%%%%%%%%

%\appendix


%If you want to present additional material which would interrupt the flow of the %main paper,

%%%%%%%%%%%%%%%%%%%%%%%%%%%%%%%%%%%%%%%%%%%%%%%%%%

% Don't change these lines
\bsp	% typesetting comment
\label{lastpage}
\end{document}